\documentclass[11pt, draftcls, onecolumn, journal]{IEEEtran}
\usepackage{amsmath,epsfig,bbm,amsbsy,cite,amssymb,amsthm}
\usepackage{cite,subfigure,psfrag,graphicx,amssymb,color}
\usepackage{graphicx,amssymb}

\begin{document}
\title{Segmented compressed sampling for analog-to-information conversion:
\\ Method and performance analysis}

\author{Omid~Taheri,~\IEEEmembership{Student Member,~IEEE,}
        and~Sergiy~A.~Vorobyov,~\IEEEmembership{Senior Member,~IEEE}
\thanks{The authors are with the Department of Electrical and Computer
Engineering, University of Alberta, Edmonton, Alberta, Canada. The
contacting emails are \{otaheri, vorobyov\}@ece.ualberta.ca.}
\thanks{{\bf Corresponding author:} Sergiy A. Vorobyov, Dept. of \
Electrical and Computer Engineering, University of Alberta,
9107-116~St., Edmonton, Alberta, T6G 2V4, Canada; Phone: +1 (780)
492 9702, Fax: +1 (780) 492 1811.}
\thanks{This work was supported in parts by research grants from the
Natural Science and Engineering Research Council (NSERC) of Canada
and Alberta Ingenuity New Faculty Award.} }

\maketitle

\begin{abstract}
A new segmented compressed sampling method for
analog-to-information conversion (AIC) is proposed. An analog
signal measured by a number of parallel branches of mixers and
integrators (BMIs), each characterized by a specific random
sampling waveform, is first segmented in time into $M$ segments.
Then the sub-samples collected on different segments and different
BMIs are reused so that a larger number of samples than the number
of BMIs is collected. This technique is shown to be equivalent to
extending the measurement matrix, which consists of the BMI
sampling waveforms, by adding new rows without actually increasing
the number of BMIs. We prove that the extended measurement matrix
satisfies the restricted isometry property with overwhelming
probability if the original measurement matrix of BMI sampling
waveforms satisfies it. We also show that the signal recovery
performance can be improved significantly if our segmented AIC is
used for sampling instead of the conventional AIC. Simulation
results verify the effectiveness of the proposed segmented
compressed sampling method and the validity of our theoretical
studies.
\end{abstract}


\IEEEpeerreviewmaketitle

\vspace{-0.7cm}
\section{Introduction}
According to Shannon's sampling theorem, an analog band-limited
signal can be recovered from its discrete-time samples if the
sampling rate is at least twice the maximum frequency present in
the signal. Recent theory of compressed sampling (CS), however,
suggests that a signal can be recovered from fewer samples if it is
sparse or compressible \cite{CandesWakin08, CandesTao05,
Donoho06b, HauptNowak06}.
CS theory also suggests that a universal sampling matrix (for
example, a random projection matrix) can be designed, and it can
be used for all sparse signals regardless of their nature
\cite{CandesTao05}. CS has already found a wide range of
applications such as image acquisition \cite{Wakinetal06}, sensor
networks \cite{Bajwaetal07a}, cognitive radios
\cite{YuHoyosSadler08}, communication channel estimation
\cite{TaubockHlawatsch08,Bajwaetal08}, etc.

The sampling process often used in the CS literature consists of
two steps. First, an analog signal is sampled at the Nyquist rate
and then a measurement matrix is applied to the time domain
samples in order to collect the compressed samples (see, for
example, \cite{YuHoyosSadler08}). This sampling approach, however,
defeats one of the primary purposes of CS, which is avoiding high
rate sampling. A more practical approach for ``direct'' sampling
and compression of analog signals has been presented in
\cite{Eldar09}. The analog signal is assumed to belong to the
class of signals in shift-invariant spaces, that is, the analog
signal can be represented as a linear combination of a set of $m$
basis functions defined over a period $T$. The analog signal is
first passed through a filter bank where each filter is matched to
one of the $m$ basis functions and the output is sampled at time
instances $nT$ where $n$ is an integer. If the signal is sparse,
then only $S < m$ samples are nonzero. The set of $m$ output
samples are then passed through a measurement matrix to create $K
\geq S$ compressed samples representing the analog signal in a
specific period $[(n-1)T, \, nT]$. It is worth mentioning that
this method is a generalization of another method in
\cite{MishaliEldar09} which is devised for sub-Nyquist sampling of
multi-band signals. The limits of this method come from the
underlying assumption that the signal belongs to the class of
signals in shift-invariant spaces. Although this assumption is
argued to be valid for a variety of engineering applications
\cite{Eldar09}, \cite{deBoor} and can be generalized to the
signals in a union of subspaces \cite{LuDo},
\cite{EldarMishali09}, it is still a limiting assumption.
Moreover, the complexity of this method is by no means lower than
the complexity of another practical approach to CS, which avoids
high rate sampling \cite{CandesWakin08}, \cite{Laskaetal08}. The
name analog-to-information converter (AIC) has been coined for the
latter method. The AIC consists of several parallel branches of
mixers and integrators (BMIs) in which the analog signal is
measured against different random sampling waveforms. Therefore,
for every collected compressed sample, there is a BMI that
multiplies the signal to a sampling waveform and then integrates
the result over a period $T$.

In this paper, we propose a new segmented AIC structure with the
goal of reducing the hardware complexity.\footnote{Some
preliminary results have been reported in \cite{OmidSergiy}.} The
contributions of this work are the following. (i)~A new segmented
AIC structure is developed. In this structure, the integration
period $T$ is divided into $M$ equal subperiods such that the
sampling rate of our segmented AIC scheme is $M$ times higher than
of the AIC of \cite{CandesWakin08}. The sub-samples collected over
different subperiods by combining the sub-samples from different
BMIs are then reused in order to build additional samples. In this
way, a number of samples larger than the number of BMIs can be
collected, although such samples will be correlated. We show that
our segmented AIC technique is equivalent to extending the
measurement matrix which consists of the BMI sampling waveforms by
adding new rows without actually increasing the number of BMIs. In
this respect, the following works also need to be mentioned
\cite{Bajwaetal07b}, \cite{BarStee07}. In \cite{Bajwaetal07b},
Toeplitz-structured measurement matrices are considered, while
measurement matrices built on one random vector with shifts of $D
\geq 1$ in between the rows appear in radar imaging application
considered in \cite{BarStee07}. (ii)~We show that the restricted
isometry property (RIP), that is a sufficient condition for signal
recovery based on compressed samples, is satisfied for the
extended measurement matrix resulting from the segmented AIC
structure with overwhelming probability if the original matrix of
BMI sampling waveforms satisfies the RIP. Thus, our segmented AIC
is a valid candidate for CS. (iii)~We also show that the signal
recovery performance improves if our segmented AIC is used for
sampling instead of the AIC of \cite{CandesWakin08} with the same
number of BMIs. The mathematical challenge in this part of the
work is that the samples collected by our segmented AIC are
correlated, while all available results on performance analysis of
the signal recovery are obtained for the case of uncorrelated
samples.

The rest of this paper is organized as follows. Necessary
background on CS, CS signal recovery, and AIC is briefly
summarized in Section~\ref{sec:prelims}. The main idea of the
paper, that is, the segmented AIC structure, is explained in
Section~\ref{sec:segmentedsamplingmethod}. We prove in
Section~\ref{sec:ripforsegmentedcs} that the extended measurement
matrix resulting from the proposed segmented AIC satisfies the RIP
and, therefore, the segmented AIC is a legitimate CS method. The
signal recovery performance analysis for our segmented AIC is
summarized in Section~\ref{sec:perfanalysis}.
Section~\ref{sec:simresults} demonstrates the simulation results
and Section~\ref{sec:conclusion} concludes the paper.

\section{Background}
\label{sec:prelims} {\it CS basics and notations:} CS deals with a
low rate representation of sparse signals, i.e., such signals
which have few nonzero projections on the vectors of an orthogonal
basis (sparsity basis). Let  $\boldsymbol{\Psi} = \left(
\boldsymbol{\psi}_{1}^T, \boldsymbol{\psi}_{2}^T, \hdots,
\boldsymbol{\psi}_{N}^T \right)^T$ be an $N \times N$ matrix of
basis vectors $\boldsymbol{\psi}_{i}, \;i=1, \hdots, N$, i.e., the
sparsity basis, and $\boldsymbol{f}$ be a discrete-time sparse
signal\footnote{It can be in $\mathbbm{R}^N$ or $\mathbbm{C}^N$.}
represented in this basis as
\begin{equation}
    \boldsymbol{f}=\sum_{i=1}^{N}{x_{i}\boldsymbol{\psi}_{i}^H} =
    \boldsymbol{\Psi}^{H} \boldsymbol{x}
    \label{eq:signalrepresentation}
\end{equation}
where $\boldsymbol{x} = \left(x_{1}, x_2, \hdots, x_N\right)^T$ is
the $N \times 1$ vector of coefficients and $(\cdot )^T$ and
$(\cdot )^H$ stand for the transpose and Hermitian transpose,
respectively. A signal is $S$-sparse if at most $S$ projections on
the rows of $\boldsymbol{\Psi}$, i.e., coefficients of
$\boldsymbol{x}$, are nonzero. It is known that a universal
compressed sampling method can be designed to effectively sample
and recover $S$-sparse signals regardless of the specific sparsity
domain \cite{CandesWakin08}, \cite{CandesTao05}.

Among various bounds on the sufficient number of collected
compressed samples\footnote{See \cite{DonohoTanner09} for broader
review.} $K$ ($S < K < N$) required for recovering an $S$-sparse
signal, the first and most popular one is given by the following
inequality $S\leq C K/\text{log}(N/K)$ where $C$ is some constant
\cite{CandesWakin08}. This bound is derived based on the uniform
uncertainty principle \cite{CandesTao06}. Let $\boldsymbol{\Phi}$
be a $K~\times~N$ measurement matrix applied to a sparse signal
for collecting $K$ compressed samples. Then the uniform
uncertainty principle states that $\boldsymbol{\Phi}$ must satisfy
the following restricted isometry property (RIP)
\cite{CandesWakin08}. Let $\boldsymbol{\Phi}_{\cal T}$ be a
sub-matrix of $\boldsymbol{\Phi}$ retaining only the columns with
their indexes in the set ${\cal T} \subset \{ 1, \hdots , N \}$.
Then the $S$-restricted isometry constant $\delta_S$ is the
smallest number satisfying the inequality
\begin{equation}
\frac{K}{N} (1 - \delta_S) \| \boldsymbol{c} \|_{l_{2}}^2 \leq \|
\boldsymbol{\Phi}_{\cal T} \boldsymbol{c} \|_{l_{2}}^2 \leq
\frac{K}{N} (1 + \delta_S ) \| \boldsymbol{c} \|_{l_{2}}^2
\label{eq:ripconstant}
\end{equation}
for all sets ${\cal T}$ of cardinality less than or equal to $S$
and all vectors $\boldsymbol{c}$ (here $\|\cdot \|_{l_2}$ denotes
the Euclidean norm of a vector). As shown in \cite{CandesTao05},
\cite{Baraniuketal08}, if the entries of $\boldsymbol{\Phi}$ are,
for example, independent zero mean Gaussian variables with
variance $1/N$, then $\boldsymbol{\Phi}$ satisfies the RIP for
$S\leq C K/\text{log}(N/K)$ with high probability.\footnote{Note
that in order to ensure consistency throughout the paper, the
variance of the elements in $\boldsymbol{\Phi}$ is taken to be
$1/N$ instead of $1/K$ as, for example, in \cite{CandesTao05}.
Thus, the multiplier $K/N$ is added in the left- and right-hand
sides of \eqref{eq:ripconstant}.}

{\it Recovery methods:} Using the measurement matrix
$\boldsymbol{\Phi}$, the $1 \times K$ vector of compressed samples
$\boldsymbol{y}$ can be calculated as
$\boldsymbol{y}~=~\boldsymbol{\Phi}
\boldsymbol{f}~=~\boldsymbol{\Phi}^{\prime} \boldsymbol{x}$ where
$\boldsymbol{\Phi}^{\prime} = \boldsymbol{\Phi}
\boldsymbol{\Psi}^{H}$. A signal can be recovered from
its noiseless sample vector $\boldsymbol{y}$ based on the
following convex optimization problem that can be solved by a
linear program \cite{CandesTao05}, \cite{Donoho06a}
\begin{equation}
    \text{min} \|\tilde{\boldsymbol{x}}\|_{l_{1}} \quad
    \text{subject to} \quad \boldsymbol{\Phi}^{'} \tilde
    {\boldsymbol{x}}=\boldsymbol{y}
    \label{eq:loneminimizationchanged}
\end{equation}
where $\| \cdot \|_{l_{1}}$ denotes the ${l_{1}}$-norm of a
vector.

If the compressed samples are noisy, the sampling process can
be expressed as
\begin{equation} \label{noisymodel}
\boldsymbol{y} = \boldsymbol{\Phi} \boldsymbol{f} + \boldsymbol{w}
\end{equation}
where $\boldsymbol{w}$ is a zero mean noise vector with
identically and independently distributed (i.i.d.) entries of
variance $\sigma^2$. Then the recovery problem is modified as
\cite{CandesRombergTao05}
\begin{equation}
    \text{min} \|\tilde{\boldsymbol{x}}\|_{l_{1}} \quad
    \text{subject to} \quad \|\boldsymbol{\Phi}^{'}
    \tilde{\boldsymbol{x}}-\boldsymbol{y}\|_{l_2}\leq \gamma
    \label{eq:noisyloneminimization}
\end{equation}
where $\gamma$ is the bound on the square root of the noise
energy.

Another technique for sparse signal recovery from noisy samples
(see \cite{HauptNowak06}) uses the empirical risk minimization
method that was first developed in statistical learning theory for
approximating an unknown function based on noisy measurements
\cite{Vapnik98}. Note that the empirical risk minimization-based
recovery method is of a particular interest since under some
simplifications (see \cite[p.~4041]{HauptNowak06}) it reduces to
another well-known least absolute shrinkage and selection operator
(LASSO) method \cite{AngelosanteGiannakis09}. Therefore, the risk
minimization-based method of \cite{HauptNowak06} provides the
generality which we need in this paper.

In application to CS, the unknown function is the sparse signal
and the noisy compressed samples are the collected data. 
Let the entries of the measurement matrix $\boldsymbol{\Phi}$ be
selected with equal probability as $\pm1/\sqrt{N}$, and the energy
of the signal $\boldsymbol{f}$ be bounded so that $\|
\boldsymbol{f} \|^2 \leq NB^2$. The risk $r(\hat{\boldsymbol{f}})$
of a candidate reconstruction $\hat{\boldsymbol{f}}$ and its
empirical risk $\hat{r}(\hat{\boldsymbol{f}})$ are defined as
follows \cite{Vapnik98}
\begin{equation}
\label{eq:riskdefs}
    r(\hat{\boldsymbol{f}}) =
    \frac{\|\hat{\boldsymbol{f}} - \boldsymbol{f} \|^2}{N}+\sigma^{2},
    \qquad
    \hat{r}(\hat{\boldsymbol{f}}) = \frac{1}{K}\sum_{j=1}^{K}{\left(y_{j}
    -\boldsymbol{\phi}_{j} \hat{\boldsymbol{f}} \right)^2}.
\end{equation}
Then the candidate reconstruction $\hat{\boldsymbol{f}}_K$
obtained based on $K$ samples can be found as
\cite{HauptNowak06}
\begin{equation}
    \hat{\boldsymbol{f}}_{K} = \arg \underset{\hat{\boldsymbol{f}} \in
    {\cal F} (B)}{\min} \left\{\hat{r}(\hat{\boldsymbol{f}}) +
    \frac{c(\hat{\boldsymbol{f}}) \log 2}{\epsilon K}\right\}
    \label{eq:empriskminequation}
\end{equation}
where ${\cal F} (B) = \{ \boldsymbol{f}: \|\boldsymbol{f}\|^2 \leq
N B^{2} \}$,
$c(\hat{\boldsymbol{f}})$ is a nonnegative number assigned to a
candidate signal $\hat{\boldsymbol{f}}$, and $\epsilon = 1/\left(
50 (B+\sigma)^2 \right)$. Moreover, $\hat{\boldsymbol{f}}_{K}$
given by \eqref{eq:empriskminequation} satisfies the following
inequality \cite{HauptNowak06}
\begin{equation}
    E \left\{\frac{\|\hat{\boldsymbol{f}}_{K} - \boldsymbol{f}
    \|^2}{N}\right\} \leq
    C_{1}\underset{\hat{\boldsymbol{f}}\in {\cal F} (B) }{\min}
    \left\{\frac{\|\hat{\boldsymbol{f}} - \boldsymbol{f} \|^2}{N} +
    \frac{c(\hat{\boldsymbol{f}})\log 2+4}{\epsilon K}\right\}
    \label{eq:empriskminerror}
\end{equation}
where $C_1 = [(27 - 4e) (B/\sigma)^{2} + (50-4\sqrt{2}) B / \sigma
+ 26] / [(23-4e) (B/\sigma)^{2} + (50-4\sqrt{2}) B / \sigma +
24]$,  $e = 2.7183 \hdots$, and  $E\{ \cdot \}$ stands for the
expectation operation.

Let a compressible signal $\boldsymbol{f}$ be defined as a signal
for which $\|\boldsymbol{f}^{(m)} - \boldsymbol{f}\|^2 \leq N
C_{A} m^{-2 \alpha}$ where $\boldsymbol{f}^{(m)}$ is the best
$m$-term approximation of $\boldsymbol{f}$ which is obtained by
retaining the $m$ most significant coefficients of vector
$\boldsymbol{x}$ ($\boldsymbol{x}$ being the representation of
$\boldsymbol{f}$ in the sparsity basis $\boldsymbol{\Psi}$), and
$C_{A} > 0$ and $\alpha \geq 0$ are some constants. Let also
${\cal F}_{c} (B, \alpha, C_{A}) = \{\boldsymbol{f}:
\|\boldsymbol{f}\|^2 \leq NB^{2}, \| \boldsymbol{f}^{(m)} -
\boldsymbol{f} \|^2 \leq N C_{A} m^{-2\alpha} \}$ be the set of
compressible signals. Then based on the weight assignment $c
(\boldsymbol{f}) = 2 \log(N) N_{\boldsymbol{x}}$ (here
$N_{\boldsymbol{x}}$ is the actual number of nonzero coefficients
in $\boldsymbol{x}$) the following inequality holds
\cite{HauptNowak06}
\begin{equation}
    \underset{\boldsymbol{f} \in {\cal F}_{c}(B,\alpha,C_{A})}{\sup}
    E\left\{\frac{\|\hat{\boldsymbol{f}}_{K} - \boldsymbol{f} \|^2}{N}
    \right\} \leq C_{1} C_{2} \left( \frac{K}{\log N} \right)^{-2
    \alpha/(2\alpha+1)}
    \label{eq:originalsupcomperror}
\end{equation}
where $C_{2}=C_{2}(B,\sigma,C_{A})>0$ is a constant.

If signal $\boldsymbol{f}$ is indeed sparse and belongs to
${\cal F}_{s}(B,S)=\{\boldsymbol{f}:\|\boldsymbol{f}\|^{2}\leq
NB^{2},\|\boldsymbol{f}\|_{l_0}\leq S\}$, then there exists a
constant $C^{\prime}_{2}=C^{\prime}_{2}(B,\sigma) > 0$ such that
\cite{HauptNowak06}
\begin{equation}
    \underset{\boldsymbol{f}\in {\cal F}_{s}(B,S)}{\sup}
    E\left\{\frac{\|\hat{\boldsymbol{f}}_{K} - \boldsymbol{f}\|^2}{N}
    \right\} \leq
    C_{1}C^{\prime}_{2} \left( \frac{K}{S\log N} \right)^{-1}.
    \label{eq:originalcomperror}
\end{equation}

\begin{figure}[ht]
\centering
\begin{picture}(0,0)%
\special{psfile=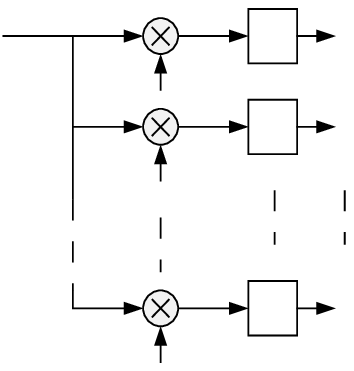}%
\end{picture}%
\setlength{\unitlength}{4144sp}%
\begingroup\makeatletter\ifx\SetFigFont\undefined
\def\x#1#2#3#4#5#6#7\relax{\def\x{#1#2#3#4#5#6}}%
\expandafter\x\fmtname xxxxxx\relax \def\y{splain}%
\ifx\x\y   
\gdef\SetFigFont#1#2#3{%
  \ifnum #1<17\tiny\else \ifnum #1<20\small\else
  \ifnum #1<24\normalsize\else \ifnum #1<29\large\else
  \ifnum #1<34\Large\else \ifnum #1<41\LARGE\else
     \huge\fi\fi\fi\fi\fi\fi
  \csname #3\endcsname}%
\else
\gdef\SetFigFont#1#2#3{\begingroup
  \count@#1\relax \ifnum 25<\count@\count@25\fi
  \def\x{\endgroup\@setsize\SetFigFont{#2pt}}%
  \expandafter\x
    \csname \romannumeral\the\count@ pt\expandafter\endcsname
    \csname @\romannumeral\the\count@ pt\endcsname
  \csname #3\endcsname}%
\fi
\fi\endgroup
\begin{picture}(1588,1781)(3702,-3953)
\put(4861,-2452){\makebox(0,0)[lb]{\smash{\SetFigFont{10}{12.0}{sf}$\int_{0}^{T}$}}}
\put(3714,-2354){\makebox(0,0)[lb]{\smash{\SetFigFont{12}{14.4}{sf}$f(t)$}}}
\put(4476,-2561){\makebox(0,0)[lb]{\smash{\SetFigFont{10}{12.0}{sf}$\Phi_{1}(t)$}}}
\put(4476,-3059){\makebox(0,0)[lb]{\smash{\SetFigFont{10}{12.0}{sf}$\Phi_{2}(t)$}}}
\put(4476,-3889){\makebox(0,0)[lb]{\smash{\SetFigFont{10}{12.0}{sf}$\Phi_{K}(t)$}}}
\put(4861,-2867){\makebox(0,0)[lb]{\smash{\SetFigFont{10}{12.0}{sf}$\int_{0}^{T}$}}}
\put(4861,-3696){\makebox(0,0)[lb]{\smash{\SetFigFont{10}{12.0}{sf}$\int_{0}^{T}$}}}
\put(5238,-2830){\makebox(0,0)[lb]{\smash{\SetFigFont{10}{12.0}{sf}$y_2$}}}
\put(5238,-3640){\makebox(0,0)[lb]{\smash{\SetFigFont{10}{12.0}{sf}$y_K$}}}
\put(5238,-2395){\makebox(0,0)[lb]{\smash{\SetFigFont{10}{12.0}{sf}$y_1$}}}
\end{picture}
\caption{The structure of the AIC based on RMPI.}
\label{fig:parallelaicstructure}
\end{figure}

{\it AIC:} The random modulation preintegration (RMPI) structure
is proposed for AIC in \cite{CandesWakin08}. The RMPI multiplies
the signal and the sampling waveforms in the analog domain and
then integrates the product over the signal period to produce
samples. It implies that the sampling device has a number of
parallel BMIs in order to process the analog signal in real-time.
The RMPI structure is shown in
Fig.~\ref{fig:parallelaicstructure}, where $f(t)$ is the analog
signal being sampled, $\boldsymbol{\phi}_{i}(t), \; i=1,\hdots,K$
are the sampling waveforms (rows of the measurement matrix
$\boldsymbol{\Phi}$), and $y_{i}, \; i=1,\hdots, K$ are the
compressed samples.

\section{Segmented Compressed Sampling Method}
\label{sec:segmentedsamplingmethod}

AIC removes the need for high speed sampling, but it may still be
necessary in many practical applications to collect a larger
number of compressed samples than the AIC hardware (the number of
parallel BMIs) may allow. Indeed, a smaller number of samples may
have a negative effect on the signal recovery accuracy which can
be an issue in a number of applications. In order to collect a
larger number of compressed samples using AIC, we need to increase
the hardware complexity by adding more BMIs. The latter makes the
AIC
device complex and expensive although its sampling rate is much
lower than that of analog-to-digital converter (ADC). 
Therefore, it is desirable to reduce the number of parallel BMIs
in AIC without sacrificing the signal recovery accuracy. It can be
achieved by adding to AIC the capability of sampling at a higher
rate, which is, however, significantly lower than the sampling
rate required by ADC. The latter can be achieved by splitting the
integration period $T$ in every BMI of the AIC in
Fig.~\ref{fig:parallelaicstructure} into shorter subperiods. It is
equivalent to generating a number of incomplete samples of a
signal. Note that since the original integration period is divided
into a number of smaller subperiods, the samples collected over
all parallel BMIs during one subperiod do not have complete
information about the signal. Therefore, they are called
incomplete samples. Hereafter, the complete samples obtained over
the whole period $T$ are referred to as just samples, while the
incomplete samples are referred to as sub-samples.

\subsection{The Basic Idea and the Model}
\label{ssec:ideaofsegmentedsampling}

The basic idea is to collect the sub-samples as described above
and then reuse them in order to build additional samples. In this
manner, a larger number of samples than the number of BMIs can be
collected. It allows for a tradeoff between AIC and ADC since as
in AIC the signal is measured at a low rate by correlating it to a
number of sampling waveforms, while the integration period is
split into shorter sub-intervals which is similar to the
requirement of a higher sampling rate as in ADC. However, the
required sampling rate in the proposed scheme is still
significantly lower than that required by ADC.

Let the integration period be split into $M$ sub-intervals, and
let $\boldsymbol{y}_{k}=$$\left(y_{k,1}, \;\hdots ,\;
y_{k,M}\right)^T, \; k=1, \hdots ,K$ be the vectors of sub-samples
collected against the sampling waveforms $\boldsymbol{\phi}_{k},
\; k=1, \hdots ,K$, where $K$ is the original number of sampling
waveforms, i.e., the number of BMIs. The sub-sample $y_{k,j}$ is
given by
\begin{equation}
y_{k,j}=\int_{(j-1)T/M}^{jT/M}{x(t)\boldsymbol{\phi}}_{k}(t) dt.
\end{equation}
Then the total number of sub-samples collected in all BMIs over
all subperiods is $M K$. These sub-samples can be gathered in the
following $K \times M$ matrix
\begin{align}
\boldsymbol{Y} =
    \left(
        \begin{array}{cccc}
    y_{1,1} & y_{1,2} & \hdots & y_{1,M}\\
    y_{2,1} & y_{2,2} & \hdots & y_{2,M}\\
    \vdots & \vdots & \vdots & \vdots\\
    y_{K,1} & y_{K,2} & \hdots & y_{K,M}\\
        \end{array}
    \right)
    \label{eq:matrixofsubsamples}
\end{align}
where the $k$-th row contains the sub-samples obtained by
correlating the measured signal with the waveform
$\boldsymbol{\phi}_k$ over $M$ subperiods each of length $T/M$.

The original $K$ samples, i.e., the samples collected at BMIs over
the whole time period $T$, are
\begin{equation} \label{eq11}
y_{k}=\sum_{m=1}^{M} [\boldsymbol{Y}]_{k,m}, \qquad k=1, \hdots, K
\end{equation}
where $[\boldsymbol{Y}]_{k,m}$ denotes the $(k,m)$-th element of
$\boldsymbol{Y}$, that is, $[\boldsymbol{Y}]_{k,m} = y_{k,m}$.

In order to construct additional samples to the samples obtained
using \eqref{eq11}, we consider columnwise permuted versions of
$\boldsymbol{Y}$. The following definitions are then in order.

The {\it permutation} $\pi$ is a one-to-one mapping of the elements
of a set ${\cal D}$ to itself by simply changing the order of the
elements. Then $\pi (k)$ stands for the index of the $k$-th
element in the permuted set. For example, let ${\cal D}$ consists
of the elements of a $K \times 1$ vector $\boldsymbol{z}$, and the
order of the elements in ${\cal D}$ is the same as in
$\boldsymbol{z}$. After applying the permutation function $\pi$ to
$\boldsymbol{z}$, the permuted vector is $\boldsymbol{z}^\pi = \left(
z_{\pi (1)}, \hdots, z_{\pi (k)}, \hdots, z_{\pi (K)} \right)^T$. If
vector $\boldsymbol{z}$ is itself the vector of indexes, i.e.,
$\boldsymbol{z} = (1, \hdots, K )^T$, then obviously $z_{\pi (k)}
= \pi (k)$.

The permuted versions of the sub-sample matrix $\boldsymbol{Y}$
can be obtained by applying different permutations to different
columns of $\boldsymbol{Y}$. Specifically, let ${\cal P}^{(i)} =
\{ \pi^{(i)}_{1}, \hdots, \pi^{(i)}_{j}, \hdots, \pi^{(i)}_{M} \}$
be the $i$-th set of column permutations with $\pi^{(i)}_{j}$
being the permutation function applied to the $j$-th column of
$\boldsymbol{Y}$, and let $I$ stand for the number of such
permutation sets. Then according to the above notations, the
matrix resulting from applying the set of permutations ${\cal
P}^{(i)}$ to the columns of $\boldsymbol{Y}$ can be expressed as
$\boldsymbol{Y}^{{\cal P}^{(i)}} =
\left(\boldsymbol{y}_1^{\pi^{(i)}_{1}}, \hdots,
\boldsymbol{y}_j^{\pi^{(i)}_{j}}, \hdots,
\boldsymbol{y}_M^{\pi^{(i)}_{M}}\right)$ where $\boldsymbol{y}_j$
is the $j$-th column of $\boldsymbol{Y}$.

Permutation sets ${\cal P}^{(i)}, \; i = 1, \hdots, I$ are chosen
in such a way that all sub-samples in a specific row of
$\boldsymbol{Y}^{{\cal P}^{(i)}}$ come from different rows of the
original sub-sample matrix $\boldsymbol{Y}$ as well as from
different rows of other permuted matrices $\boldsymbol{Y}^{{\cal
P}^{(1)}}, \hdots, \boldsymbol{Y}^{{\cal P}^{(i-1)}}$. For
example, all sub-samples in a specific row of
$\boldsymbol{Y}^{{\cal P}^{(1)}}$ must come from different rows of
the original matrix $\boldsymbol{Y}$ only, while the sub-samples
in a specific row of $\boldsymbol{Y}^{{\cal P}^{(2)}}$ come from
different rows of $\boldsymbol{Y}$ and $\boldsymbol{Y}^{{\cal
P}^{(1)}}$ and so on. This requirement is forced to make sure that
any additional sample has the least possible correlation with the
original samples of \eqref{eq11}. Then the additional $K \, I$
samples can be obtained based on the permuted matrices
$\boldsymbol{Y}^{{\cal P}^{(i)}}, \; i=1, \hdots, I$ as
\begin{equation}
y_{k}^{{\cal P}^{(i)}}=\sum_{m=1}^{M} [\boldsymbol{Y}^{{\cal
P}^{(i)}}]_{k,m}, \qquad k=1,\hdots,K \quad i=1, \hdots, I .
\label{eq:firstsetofnewsamples}
\end{equation}

It is worth noting that in terms of the hardware structure, the
sub-samples used to generate additional samples must be chosen
from different BMIs as well as different integration subperiods.
This is equivalent to collecting additional samples by correlating
the signal with additional sampling waveforms which are not
present among the actual BMI sampling waveforms. Each of these
additional sampling waveforms comprises the non-overlapping
subperiods of $M$ different original waveforms.

Now the question is how many permuted matrices, which satisfy the
above summarized conditions, can be generated based on
$\boldsymbol{Y}$. Consider the following $K \times M$ matrix
\begin{equation}
\boldsymbol{Z} \triangleq \underbrace{(\boldsymbol{z},
\boldsymbol{z}, \hdots, \boldsymbol{z})}_{M \; \text{times}}
\label{eq:definitionofmatrixx}
\end{equation}
where $\boldsymbol{z}$ is the vector
of indexes. Applying the column permutation set ${\cal P}^{(i)}$
to the columns of $\boldsymbol{Z}$, we obtain a permuted matrix
$\boldsymbol{Z}^{{\cal P}^{(i)}} =
\left(\boldsymbol{z}^{\pi^{(i)}_{1}}, \hdots,
\boldsymbol{z}^{\pi^{(i)}_{j}}, \hdots,
\boldsymbol{z}^{\pi^{(i)}_{M}}\right)$. Then the set of all permuted
versions of $\boldsymbol{Z}$ can be denoted as ${\cal
S}_{\boldsymbol{Z}} = \{ \boldsymbol{Z}^{{\cal P}^{(1)}}, \hdots,
\boldsymbol{Z}^{{\cal P}^{(I)}} \}$. With these notations, the
following theorem is in order.

\newtheorem{kminus1newsets}{Theorem}
\begin{kminus1newsets}
\label{thm:kminus1newsets} The size of ${\cal
S}_{\boldsymbol{Z}}$, i.e., the number $I$ of permutation sets
${\cal P}^{(i)}, \; i = 1, \hdots, I$ which satisfy the conditions
\begin{eqnarray} \label{seq:requirementonentriesofA}
& &\!\!\!\!\!\!\!\!\!\!\!\!\!\!\! [\boldsymbol{Z}^{{\cal
P}^{(i)}}]_{k,j} \neq [\boldsymbol{Z} ^{{\cal P}^{(i)}}]_{k,r}, \;
\forall \boldsymbol{Z}^{{\cal P}^{(i)}} \in {\cal
S}_{\boldsymbol{Z}}, \;\; j\neq r, \; k \in \{ 1, \hdots,
K \}, \; j,r \in \{ 1, \hdots, M \} \\
& &\!\!\!\!\!\!\!\!\!\!\!\!\!\!\! \exists! j \; \text{or} \;
\nexists j \; \text{such that} \; [\boldsymbol{Z}^{{\cal
P}^{(i)}}]_{k,j} = [\boldsymbol{Z}^{{\cal P}^{(l)}}]_{h,j}, \quad
\forall \boldsymbol{Z}^{{\cal P}^{(i)}}, \boldsymbol{Z}^{{\cal
P}^{(l)}} \in {\cal S}_{\boldsymbol{Z}}, \; \boldsymbol{Z}^{{\cal
P}^{(i)}} \neq \boldsymbol{Z}^{{\cal P}^{(l)}}, \forall j \in \{1,
\hdots,
M\} \nonumber \\
& & \forall k,h \in \{1, \hdots, K\}
\label{seq:requirementonsxelements}
\end{eqnarray}
is at most $K-1$. Here $[\boldsymbol{Z}^{{\cal P}^{(i)}}]_{k,j}$
stands for the $(k,j)$-th element of the permuted matrix
$\boldsymbol{Z}^{{\cal P}^{(i)}}$.
\end{kminus1newsets}

\newtheorem{remarkcndsonperms}{Remark}
\begin{remarkcndsonperms}
\label{rmrk:remarkcndsonperms} Using the property that $z_{\pi
(k)} = \pi (k)$ for the vector of indexes $\boldsymbol{z}$, the
conditions \eqref{seq:requirementonentriesofA} and
\eqref{seq:requirementonsxelements} can also be expressed in terms
of permutations as
\begin{align}
\label{seq:requirementonpermi} & \pi^{(i)}_j (k) \neq
\pi^{(i)}_r (k) \quad \forall i\in\{1,\hdots,I\}, \; j \neq r, \;
k \in \{1,\hdots,K\}, \; j,r \in \{1,\hdots,M\} \\
& \exists! j \; \text{or} \; \nexists j \; \text{such that} \;
\pi^{(i)}_j (k) \! = \!  \pi^{(l)}_j (h) \; \forall i, l \!\in\!
\{1, \hdots, I\}, \;\! i \!\neq \!l, \; \forall j \!\in \!\{1,
\hdots, M\}, \forall k, h \!\in \!\{1, \hdots, K\} .
\label{seq:requirementonpermiandj} \\
\nonumber
\end{align}
\end{remarkcndsonperms}

\begin{proof}
See Appendix~A.
\end{proof}

{\bf Example 1:} Let the specific choice of index permutations be
$\pi_s (k) = \left( (s+k-2) \; \text{mod} \; K \right) + 1,\; s,k
= 1, \hdots, K$ with $\pi_{1}$ being the identity permutation and
'$\text{mod}$' standing for the modulo operation. For this
specific choice, $\pi^{(i)}_{j} = \pi_{[i(j-1) \; \text{mod} \; K]
+ 1 }, \; i=1, \hdots, K-1, \; j=1, \hdots, M$. Consider the
following matrix notation for the set ${\cal P}$ where the
elements along the $i$-th row are the permutations ${\cal
P}^{(i)}$, $i = 1, \hdots, I$

\begin{eqnarray}
    {\cal P} \!\!&\triangleq&\!\!
    \left( \!\!
    \begin{array}{c}
    {\cal P}^{(1)}\\
    {\cal P}^{(2)}\\
    {\cal P}^{(3)}\\
    \vdots    \\
    {\cal P}^{(K-2)}\\
    {\cal P}^{(K-1)}\\
    \end{array}
    \!\!\right)
\!\!=\!\!   \left( \!\!
    \begin{array}{ccccc}
    \pi^{(1)}_{1} & \pi^{(1)}_{2} & \pi^{(1)}_{3} &\hdots &\pi^{(1)}_{M}\\
    \pi^{(2)}_{1} & \pi^{(2)}_{2} & \pi^{(2)}_{3} &\hdots &\pi^{(2)}_{M}\\
    \pi^{(3)}_{1} & \pi^{(3)}_{2} & \pi^{(3)}_{3} &\hdots &\pi^{(3)}_{M}\\
    \vdots    & \vdots    & \vdots    &\vdots & \vdots\\
    \pi^{(K-2)}_{1} & \pi^{(K-2)}_{2} & \pi^{(K-2)}_{3} &\hdots &\pi^{(K-2)}_{M}\\
    \pi^{(K-1)}_{1} & \pi^{(K-1)}_{2} & \pi^{(K-1)}_{3} &\hdots &\pi^{(K-1)}_{M}\\
    \end{array}
    \!\!\right) \nonumber \\
\!\!&=&\!\!
    \left(\!\!
    \begin{array}{ccccc}
    \pi_{1} & \pi_{2} & \pi_{3} & \hdots &\pi_{M}\\
    \pi_{1} & \pi_{3} & \pi_{5}     & \hdots &\pi_{[2(M-1) \; \text{mod}
\; K] + 1 }\\
    \pi_{1} & \pi_{4} & \pi_{7}     & \hdots &\pi_{[3(M-1) \; \text{mod}
\; K] + 1 }\\
    \vdots    & \vdots    & \vdots  &\vdots & \vdots\\
    \pi_{1} & \pi_{K-1} & \pi_{K-3} & \hdots &\pi_{[(K-2)(M-1) \; \text{mod}
\; K] + 1 }\\
    \pi_{1} & \pi_{K}   & \pi_{K-1} & \hdots &\pi_{[(K-1)(M-1) \; \text{mod}
\; K] + 1 } \\
    \end{array}
    \!\!\right).
   \label{eq:permutationsallinamatrix}
\end{eqnarray}
Note that not all permutations ${\cal P}^{(i)}$, $i = 1, \hdots,
I$ used in \eqref{eq:permutationsallinamatrix} may be permissible.
In fact, the set of permutations ${\cal P}^{(i)}$ with $K /
gcd(i,K) < M$ has at least one repeated permutation that
contradicts the condition \eqref{seq:requirementonpermi}. Here
$gcd (\cdot, \cdot)$ stands for the greatest common devisor of two
numbers. For example, for $K=8$ and $M=4$, $K / gcd(4,K) = 2 < M$
and ${\cal P}^{(4)}$ is impermissible. Therefore, instead of
$K-1=7$, only the following $6$ sets of permutations are allowed
\begin{align}
    {\cal P}=
    \left(
    \begin{array}{cccc}
    \pi^{(1)}_{1} & \pi^{(1)}_{2} & \pi^{(1)}_{3} &\pi^{(1)}_{4}\\
    \pi^{(2)}_{1} & \pi^{(2)}_{2} & \pi^{(2)}_{3} &\pi^{(2)}_{4}\\
    \pi^{(3)}_{1} & \pi^{(3)}_{2} & \pi^{(3)}_{3} &\pi^{(3)}_{4}\\
    \pi^{(4)}_{1} & \pi^{(4)}_{2} & \pi^{(4)}_{3} &\pi^{(4)}_{4}\\
    \pi^{(5)}_{1} & \pi^{(5)}_{2} & \pi^{(5)}_{3} &\pi^{(5)}_{4}\\
    \pi^{(6)}_{1} & \pi^{(6)}_{2} & \pi^{(6)}_{3} &\pi^{(6)}_{4}\\
    \end{array}
    \right)
=
    \left(
    \begin{array}{cccc}
    \pi_{1} & \pi_{2} & \pi_{3} & \pi_{4}\\
    \pi_{1} & \pi_{3} & \pi_{5} & \pi_{7} \\
    \pi_{1} & \pi_{4} & \pi_{7} & \pi_{2} \\
    \pi_{1} & \pi_{6} & \pi_{3} & \pi_{8}\\
    \pi_{1} & \pi_{7} & \pi_{5} & \pi_{3} \\
    \pi_{1} & \pi_{8} & \pi_{7} & \pi_{6} \\
    \end{array}
    \right) .
   \label{eq:permutationsallinamatrixexaple}
\end{align}

Theorem~\ref{thm:kminus1newsets} shows how many different permuted
versions of the original sub-sample matrix $\boldsymbol{Y}$ can be
obtained such that the correlation between the original and
additional samples would be minimal. Indeed, since the set of
sub-samples that are used to build additional samples is chosen in
a way that additional samples have at most one sub-sample in
common with the previous samples, i.e., conditions
\eqref{seq:requirementonpermi} and
\eqref{seq:requirementonpermiandj} are satisfied, the set of
permutations (\ref{eq:permutationsallinamatrix}) is a valid
candidate. The $i$-th element of ${\cal P}$, i.e., the element
${\cal P}^{(i)} = \left( \pi^{(i)}_{1}, \hdots, \pi^{(i)}_{M}
\right)$, is the set of permutations applied to $\boldsymbol{Y}$
to obtain $\boldsymbol{Y}^{{\cal P}^{(i)}}$. Adding up the entries
along the rows of $\boldsymbol{Y}^{{\cal P}^{(i)}}$, a set of $K$
additional
samples can be obtained. 

\begin{figure}[ht]
\centering
\begin{picture}(0,0)%
\special{psfile=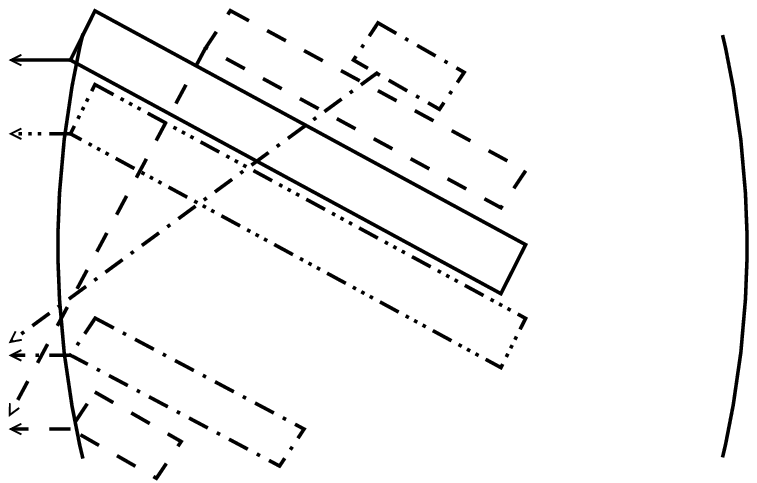}%
\end{picture}%
\setlength{\unitlength}{4144sp}%
\begingroup\makeatletter\ifx\SetFigFont\undefined
\def\x#1#2#3#4#5#6#7\relax{\def\x{#1#2#3#4#5#6}}%
\expandafter\x\fmtname xxxxxx\relax \def\y{splain}%
\ifx\x\y   
\gdef\SetFigFont#1#2#3{%
  \ifnum #1<17\tiny\else \ifnum #1<20\small\else
  \ifnum #1<24\normalsize\else \ifnum #1<29\large\else
  \ifnum #1<34\Large\else \ifnum #1<41\LARGE\else
     \huge\fi\fi\fi\fi\fi\fi
  \csname #3\endcsname}%
\else
\gdef\SetFigFont#1#2#3{\begingroup
  \count@#1\relax \ifnum 25<\count@\count@25\fi
  \def\x{\endgroup\@setsize\SetFigFont{#2pt}}%
  \expandafter\x
    \csname \romannumeral\the\count@ pt\expandafter\endcsname
    \csname @\romannumeral\the\count@ pt\endcsname
  \csname #3\endcsname}%
\fi
\fi\endgroup
\begin{picture}(3752,2182)(3858,-3851)
\put(3858,-3238){\makebox(0,0)[lb]{\smash{\SetFigFont{10}{12.0}{sf}$y_{2\!K\!-\!1}$}}}
\put(4590,-1888){\makebox(0,0)[lb]{\smash{\SetFigFont{10}{12.0}{sf}$y_{1,1}$}}}
\put(4590,-2225){\makebox(0,0)[lb]{\smash{\SetFigFont{10}{12.0}{sf}$y_{2,1}$}}}
\put(4590,-2563){\makebox(0,0)[lb]{\smash{\SetFigFont{10}{12.0}{sf}$y_{3,1}$}}}
\put(4590,-3294){\makebox(0,0)[lb]{\smash{\SetFigFont{10}{12.0}{sf}$y_{\!K\!-\!1,1}$}}}
\put(4590,-3631){\makebox(0,0)[lb]{\smash{\SetFigFont{10}{12.0}{sf}$y_{K,1}$}}}
\put(4642,-2950){\makebox(0,0)[lb]{\smash{\SetFigFont{10}{12.0}{sf}$\vdots$}}}
\put(5261,-2950){\makebox(0,0)[lb]{\smash{\SetFigFont{10}{12.0}{sf}$\vdots$}}}
\put(5208,-1888){\makebox(0,0)[lb]{\smash{\SetFigFont{10}{12.0}{sf}$y_{1,2}$}}}
\put(5208,-2225){\makebox(0,0)[lb]{\smash{\SetFigFont{10}{12.0}{sf}$y_{2,2}$}}}
\put(5209,-2563){\makebox(0,0)[lb]{\smash{\SetFigFont{10}{12.0}{sf}$y_{3,2}$}}}
\put(5209,-3293){\makebox(0,0)[lb]{\smash{\SetFigFont{10}{12.0}{sf}$y_{\!K\!-\!1,2}$}}}
\put(5209,-3631){\makebox(0,0)[lb]{\smash{\SetFigFont{10}{12.0}{sf}$y_{K,2}$}}}
\put(5936,-2950){\makebox(0,0)[lb]{\smash{\SetFigFont{10}{12.0}{sf}$\vdots$}}}
\put(5883,-1888){\makebox(0,0)[lb]{\smash{\SetFigFont{10}{12.0}{sf}$y_{1,3}$}}}
\put(5883,-2225){\makebox(0,0)[lb]{\smash{\SetFigFont{10}{12.0}{sf}$y_{2,3}$}}}
\put(5884,-2563){\makebox(0,0)[lb]{\smash{\SetFigFont{10}{12.0}{sf}$y_{3,3}$}}}
\put(5885,-3293){\makebox(0,0)[lb]{\smash{\SetFigFont{10}{12.0}{sf}$y_{\!K\!-\!1,3}$}}}
\put(5885,-3631){\makebox(0,0)[lb]{\smash{\SetFigFont{10}{12.0}{sf}$y_{K,3}$}}}
\put(6611,-2950){\makebox(0,0)[lb]{\smash{\SetFigFont{10}{12.0}{sf}$\vdots$}}}
\put(7230,-2950){\makebox(0,0)[lb]{\smash{\SetFigFont{10}{12.0}{sf}$\vdots$}}}
\put(6558,-1944){\makebox(0,0)[lb]{\smash{\SetFigFont{10}{12.0}{sf}$\cdots$}}}
\put(6558,-2281){\makebox(0,0)[lb]{\smash{\SetFigFont{10}{12.0}{sf}$\cdots$}}}
\put(6559,-2619){\makebox(0,0)[lb]{\smash{\SetFigFont{10}{12.0}{sf}$\cdots$}}}
\put(6560,-3350){\makebox(0,0)[lb]{\smash{\SetFigFont{10}{12.0}{sf}$\cdots$}}}
\put(6560,-3688){\makebox(0,0)[lb]{\smash{\SetFigFont{10}{12.0}{sf}$\cdots$}}}
\put(7050,-1888){\makebox(0,0)[lb]{\smash{\SetFigFont{10}{12.0}{sf}$y_{1,M}$}}}
\put(7050,-2225){\makebox(0,0)[lb]{\smash{\SetFigFont{10}{12.0}{sf}$y_{2,M}$}}}
\put(7050,-2563){\makebox(0,0)[lb]{\smash{\SetFigFont{10}{12.0}{sf}$y_{3,M}$}}}
\put(7050,-3293){\makebox(0,0)[lb]{\smash{\SetFigFont{10}{12.0}{sf}$y_{\!K\!-\!1,M}$}}}
\put(7050,-3631){\makebox(0,0)[lb]{\smash{\SetFigFont{10}{12.0}{sf}$y_{K,M}$}}}
\put(3858,-1888){\makebox(0,0)[lb]{\smash{\SetFigFont{10}{12.0}{sf}$y_{K+1}$}}}
\put(3859,-2225){\makebox(0,0)[lb]{\smash{\SetFigFont{10}{12.0}{sf}$y_{K+2}$}}}
\put(3915,-3632){\makebox(0,0)[lb]{\smash{\SetFigFont{10}{12.0}{sf}$y_{2K}$}}}
\end{picture}
\caption{Sub-sample selection principle for building additional
samples in Example~2.}
\label{fig:diagsampling}
\end{figure}

{\bf Example 2:} Let the number of new samples $K_a$ be at most
$K$. This means that all permutations are given by only ${\cal
P}^{(1)}$ in (\ref{eq:permutationsallinamatrix}). In this special
case, the sub-sample selection method can be summarized as
follows. For constructing the $(K+1)$-st sample, $M$ sub-samples
on the main diagonal of $\boldsymbol{Y}$ are summed up together.
Then the $M$ sub-samples on the second diagonal are used to
construct the $(K\nolinebreak+\nolinebreak2)$-nd sample, and so on
up to the $K_a$-th sample. Mathematically, the so constructed additional
samples can be expressed in terms of the elements of $\boldsymbol{Y}$ as
\begin{equation}
    y_{K+k}=\sum_{m=1}^{M}{y_{l,m}}, \qquad k=1 \hdots, K_{a}
    \label{eq:newsampleproduction}
\end{equation}
where $l=[(k+m-2)\; \text{mod}\; K]+1$ and $K_{a} \leq K$.
Fig.~\ref{fig:diagsampling} shows schematically how the
sub-samples are selected in this example.

Our segmented sampling process can be equivalently expressed in
terms of the measurement matrix. Let $\boldsymbol{\Phi}$ be the
original $K \times N$ measurement matrix. Let the $k$-th row of
the matrix $\boldsymbol{\Phi}$ be $\boldsymbol{\phi}_{k} = \left
(\boldsymbol{\phi}_{k,1}, \hdots, \boldsymbol{\phi}_{k,M} \right)$
where $\boldsymbol{\phi}_{k,j}, \; j=1, \hdots, M$ are some
vectors. Let for simplicity, the length of
$\boldsymbol{\phi}_{k,j}$ be $N/M$ and $N/M$ be an integer number.
The set of permutations applied to $\boldsymbol{Y}$ in order to
obtain $\boldsymbol{Y}^{{\cal P}^{(i)}}$ is ${\cal P}^{(i)}$. Then
the operation $\boldsymbol{\Phi}^{{\cal P}^{(i)}}$ can be
expressed as follows. The first $N/M$ columns of
$\boldsymbol{\Phi}$, which are the vectors
$\boldsymbol{\phi}_{k,1}, \; k\in\{1,...,K\}$, are permuted with
$\pi^{(i)}_{1}$. The second $N/M$ columns of $\boldsymbol{\Phi}$
are permuted with $\pi^{(i)}_{2}$ and so on until the last $N/M$
columns of $\boldsymbol{\Phi}$ which are permuted with
$\pi^{(i)}_{M}$. Then the extended measurement matrix which
combines all possible permutations ${\cal P}^{(i)}, \; i= 1,
\hdots, I$ can be expressed as
\begin{equation}
\boldsymbol{\Phi}_{e} = \left( \boldsymbol{\Phi}^T,
(\boldsymbol{\Phi}^{{\cal P}^{(1)}})^T, \hdots,
(\boldsymbol{\Phi}^{{\cal P}^{(I)}})^T \right)^T
\label{eq:largestextendedmeasurementmatrix}
\end{equation}
where $K_{e} \triangleq K + K_a = K + K I$.

{\bf Example 3:} Continuing with the set up used in Example~2, let
$K_{a} \leq K$. Then the extended measurement matrix is
\begin{equation}
\boldsymbol{\Phi}_e =
    \left(
        \begin{array}{l}
    \boldsymbol{\Phi} \\
    \boldsymbol{\Phi}_1 \\
        \end{array}
    \right)
    =
    \left(
        \begin{array}{cccc}
    \boldsymbol{\phi}_{1,1} & \boldsymbol{\phi}_{1,2} &
    \hdots & \boldsymbol{\phi}_{1,M}\\
    \vdots & \vdots & \vdots & \vdots\\
    \boldsymbol{\phi}_{K,1} & \boldsymbol{\phi}_{K,2} &
    \hdots & \boldsymbol{\phi}_{K,M}\\
    \boldsymbol{\phi}_{1,1} & \boldsymbol{\phi}_{2,2} &
    \hdots &
    \boldsymbol{\phi}_{M,M}\\
    \vdots & \vdots & \vdots & \vdots\\
    \boldsymbol{\phi}_{K_{a},1} & \boldsymbol{\phi}_{\pi_2(K_{a}),M} & \hdots &
    \boldsymbol{\phi}_{\pi_M(K_{a}),M}\\
        \end{array}
    \right)
    \label{eq:newmeasurementmatrix}
\end{equation}
where $\boldsymbol{\Phi}_1$ contains only $K_a$ rows of
$\boldsymbol{\Phi}^{{\cal P}^{(1)}}$ and $\boldsymbol{\Phi}_1 =
\boldsymbol{\Phi}^{{\cal P}^{(1)}}$ if $K_a = K$.

\subsection{Implementation Issues and Discussion}
\label{ssec:implementationconsiderations} Due to the special
structure of the extended measurement matrix
$\boldsymbol{\Phi}_{e}$, the sampling hardware needs only $K$
parallel BMIs for collecting $KI$ samples. These BMIs are
essentially the same as those in
Fig.~\ref{fig:parallelaicstructure}. The only difference is that
the integration period $T$ is divided into $M$ equal subperiods.
After every subperiod, each integrator's output is sampled and the
integrator is reset.
In addition, a multiplexer which selects the sub-samples for
constructing additional samples is needed. Note that partial sums
can be kept for constructing the samples (original and
additional), that is, the results of the integration are updated
and accumulated for each sample iteratively after each subperiod.
In this way, there is no need of designing the circuitry to
memorize the matrix of sub-samples $\boldsymbol{Y}$, but only the
partial sums for each sample are memorized at any current
subperiod.

Since the proposed segmented AIC scheme collects the sub-samples
at the $M$ times higher rate than the AIC in
Fig.~\ref{fig:parallelaicstructure}, an improved signal recovery
performance is expected. It agrees with the convention that the
recovery performance cannot be improved only due to the post
processing. Moreover, note that since the original random sampling
waveforms are linearly independent with high probability, the
additional sampling waveforms of our segmented compressed sampling
method are also linearly independent with overwhelming
probability. However, a sufficient condition that guarantees that
the extended measurement matrix of the proposed segmented AIC
scheme is an eligible choice is the RIP. Therefore, the RIP for
the proposed segmented compressed sampling scheme is analyzed in
the next section.

\section{RIP for the segmented compressed sampling method}
\label{sec:ripforsegmentedcs} The purpose of this section is to
show that the extended measurement matrix $\boldsymbol{\Phi}_{e}$
in (\ref{eq:largestextendedmeasurementmatrix}) satisfies the RIP
if the original measurement matrix $\boldsymbol{\Phi}$ satisfies
it. The latter will also imply that $\boldsymbol{\Phi}_{e}$ can be
used as a valid CS measurement matrix.
In our set up it is only assumed that the elements of the original
measurement matrix are i.i.d. zero mean Gaussian variables and the
measurement matrix is extended by adding its permuted versions as
described in the previous section.

Let us first consider the special case of Example~3. In this case,
$\boldsymbol{\Phi}$, $\boldsymbol{\Phi}_1$, and
$\boldsymbol{\Phi}_{e}$ are the original measurement matrix, the
matrix of additional sampling waveforms, and the extended
measurement matrix given by (\ref{eq:newmeasurementmatrix}),
respectively. Let the matrix $\boldsymbol{\Phi}$ satisfy the RIP
with sufficiently high probability. For example, let the elements
of $\boldsymbol{\Phi}$ be i.i.d. zero mean Gaussian random
variables with variance $1/N$. Let $\cal T$ be any subset of size
$S$ of the set $\{1, \hdots, N\}$. Then for any $0< \delta_S < 1$,
the matrix $\boldsymbol{\Phi}_{\cal T}$, which is a sub-matrix of
$\boldsymbol{\Phi}$ which consists of only the columns with their
indexes in the set ${\cal T}$ satisfies (\ref{eq:ripconstant})
with the following probability \cite{Baraniuketal08}
\begin{equation}
    \text{Pr}\{\boldsymbol{\Phi}_{\cal T} \;
    \text{satisfies (\ref{eq:ripconstant})}\} \geq 1-2
    \left(12/\delta_{S} \right)^S e^{-C_{0}\left
    (\delta_{S}/2 \right) K}
    \label{eq:phi1ripsatisfactionprob}
\end{equation}
where $C_{0} \left( \delta_{S} / 2
\right) = \delta_{S}^{2} / 16 - \delta_{S}^3 / 48$. Hereafter, the
notation $C_0$ is used instead of $C_{0} \left( \delta_{S} / 2
\right)$ for brevity.

First, the following auxiliary result on the extended measurement
matrix $\boldsymbol{\Phi}_e$ is of interest.

\newtheorem{lemripprob}{Lemma}
\begin{lemripprob}
\label{lem:lemripprob} Let the elements of the measurement matrix
$\boldsymbol{\Phi}$ be i.i.d. zero mean Gaussian variables with
variance $1/N$, $\boldsymbol{\Phi}_e$ be formed as shown in
\eqref{eq:newmeasurementmatrix}, and ${\cal T} \subset \{ 1,
\hdots, N \}$ of size $S$. If $K_a$ is chosen such that
$\text{min} \{ K, K_{a} + M - 1 \} \leq \lceil \left( K + K_{a}
\right) / 2 \rceil$, then for any $0 < \delta_S < 1$, the
following inequality holds
\begin{equation}
    \text{Pr}\{\left(\boldsymbol{\Phi}_{e}\right)_{\cal T} \;
    \text{satisfies \eqref{eq:ripconstant}}\} \geq  1-4\left(
    12 / \delta_{S} \right)^S e^{-C_{0} \lfloor
    \frac{K+K_{a}}{2}\rfloor}
    \label{eq:phiripsatisfactionprob}
\end{equation}
where $\lceil x \rceil$ and $\lfloor x \rfloor$ are the smallest
integer larger than or equal to $x$ and the largest integer
smaller than or equal to $x$, respectively, and $C_0$ is a
constant given after \eqref{eq:phi1ripsatisfactionprob}.
\end{lemripprob}

\begin{proof}
See Appendix~B.
\end{proof}

Using the above lemma, the following main result, which states
that the extended measurement matrix $\boldsymbol{\Phi}_e$ in
(\ref{eq:newmeasurementmatrix}) satisfies the RIP, can be also
proved.

\newtheorem{thmripprob}[kminus1newsets]{Theorem}
\begin{thmripprob}
\label{thm:thmripprob} Let $\boldsymbol{\Phi}_e$ be formed as in
\eqref{eq:newmeasurementmatrix} and let the elements of
$\boldsymbol{\Phi}$ be i.i.d. zero mean Gaussian variables with
variance $1/N$. If $\text{min} \{ K, K_{a} + M - 1 \} \leq \lceil
(K + K_{a}) / 2 \rceil$, then for any $0 < \delta_S < 1$, there
exist constants $C_3$ and $C_4$, which depend only on $\delta_S$,
such that for $S \leq C_{3} \lfloor (K + K_{a}) / 2 \rfloor /
\log(N / S)$ the inequality \eqref{eq:ripconstant} holds for all
$S$-sparse vectors with probability that satisfies the following
inequality
\begin{equation} \label{finprob}
    \text{Pr}\{\boldsymbol{\Phi}_{e} \;\text{satisfies RIP}\}
    \geq 1 - 4 e^{-C_{4} \lfloor (K + K_{a})/2 \rfloor}
\end{equation}
where $C_{4}=C_{0} - C_{3} \left[ 1 + \left( 1 + \log \left( 12 /
\delta_S \right) \right) / \log \left( N/S \right) \right]$ and
$C_3$ is small enough that guarantees that $C_4$ is positive.
\end{thmripprob}

\begin{proof}
See Appendix~C.
\end{proof}

Let us consider now the general case when the number of additional
samples $K_{a}$ is larger than the number of BMIs $K$, i.e.,
$K_{a} > K$, $K_{e} > 2K$, and the extended measurement matrix is
given by \eqref{eq:largestextendedmeasurementmatrix}. Note that
while proving Lemma~\ref{lem:lemripprob} for the special case of
Example~3, we were able to split the rows of $\boldsymbol{\Phi}_e$
into two sets each consisting of independent entries. In the
general case, some of the entries of the original measurement
matrix appear more than twice in the extended measurement matrix
$\boldsymbol{\Phi}_e$, and it is no longer possible to split the
rows of $\boldsymbol{\Phi}_e$ into only two sets with independent
entries. Due to the way that the additional samples are built, the
samples $y_{lK+1}, y_{lK+2}, \hdots, y_{(l+1)K}$ obtained based on
the permuted matrix $\boldsymbol{Y}^{{\cal P}^{(l)}}$, i.e., the
$l$-th set of additional samples, are uncorrelated with each
other, but they are correlated with every other set of samples
based on the original matrix $\boldsymbol{Y}$ and the permuted
matrices $\boldsymbol{Y}^{{\cal P}^{(i)}}$, $\forall i, \; i \neq
l$. Thus, the following principle can be used while partitioning
the rows of $\boldsymbol{\Phi}_e$ into the sets with independent
entries. First, the rows corresponding to the original samples
form a single set with independent entries, then the rows
corresponding to the first set of additional samples based on the
matrix $\boldsymbol{Y}^{{\cal P}^{(1)}}$ form another set and so
on. Then the number of such sets is $n_{p} = \lceil{K_{e}/K}
\rceil$, while the size of each set is
\begin{equation}
K_{i}=\left\{
\begin{array}{ll}
K, & 1 \leq i < \lceil{\frac{K_{e}}{K}}\rceil - 1 \\
K_e - (\lceil{\frac{K_e}{K}} \rceil - 1)K, & i =
\lceil{\frac{K_e}{K}} \rceil \\
\end{array}
\right. \label{eq:numofpartsgeneral}
\end{equation}
The extended measurement matrix
\eqref{eq:largestextendedmeasurementmatrix} can be rewritten as
\begin{align}
\boldsymbol{\Phi}_e = \left( \left(\boldsymbol{\Phi}_e\right)_1^T,
\left(\boldsymbol{\Phi}_e\right)_2^T, \hdots,
\left(\boldsymbol{\Phi}_e\right)_{n_{p}}^T
\right)^T \label{eq:extendedmeasurementmatrixgeneral}
\end{align}
where $\left(\boldsymbol{\Phi}_e\right)_i$ is the $i$-th partition
of $\boldsymbol{\Phi}_e$ of size given by
\eqref{eq:numofpartsgeneral}. Then the general form of
Lemma~\ref{lem:lemripprob} is as follows.

\newtheorem{lemripprobgeneral}[lemripprob]{Lemma}
\begin{lemripprobgeneral}
\label{lem:lemripprobgeneral} Let the elements of the measurement
matrix $\boldsymbol{\Phi}$ be i.i.d. zero mean Gaussian variables
with variance $1/N$, $\boldsymbol{\Phi}_e$ be the extended
measurement matrix \eqref{eq:largestextendedmeasurementmatrix}, and
${\cal T} \subset \{1, \hdots, N \}$ of size $S$. Let also $K_{a}
> K$ and $n_{p} = \lceil{K_e/K} \rceil$. Then, for any $0 <
\delta_S < 1$, the following inequality holds
\begin{equation}
\text{Pr} \{ \left( \boldsymbol{\Phi}_{e} \right)_{\cal T} \;
\text{satisfies \eqref{eq:ripconstant}} \} \geq 1-2(n_{p} - 1)
\left( 12 / \delta_{S} \right)^S \left( e^{-C_{0} K} \right) - 2
\left( 12/\delta_{S} \right)^S \left( e^{-C_{0} K_{n_{p}}} \right)
\label{eq:phigeneralripsatisfactionprob}
\end{equation}
where $K_{n_p} = K_e - \left( \lceil{ \frac{K_e}{K}} \rceil - 1
\right) K$ and $C_0$ is a constant given after
\eqref{eq:phi1ripsatisfactionprob}.
\end{lemripprobgeneral}

\begin{proof}
See Appendix~D.
\end{proof}

Lemma~2 is needed to prove that the extended measurement matrix
\eqref{eq:extendedmeasurementmatrixgeneral} satisfies the RIP.
Therefore, the general version of Theorem~\ref{thm:thmripprob} is
as follows.

\newtheorem{thmripprobgeneral}[kminus1newsets]{Theorem}
\begin{thmripprobgeneral}
\label{thm:thmripprobgeneral} Let the elements of
$\boldsymbol{\Phi}$ be i.i.d. zero mean Gaussian variables with
variance $1/N$ and $\boldsymbol{\Phi}_e$ be formed as in
\eqref{eq:largestextendedmeasurementmatrix}. If $K_{a} > K$, then
for any $0 < \delta_S < 1$, there exist constants $C_3$, $C_4$ and
$C_4^{\prime}$, such that for $S \leq C_{3} K_{n_p} / \log(N/S)$
the inequality \eqref{eq:ripconstant} holds for all $S$-sparse
vectors with probability that satisfies the following inequality
\begin{align}
\label{eq:finprobgeneral} \text{Pr} \{ \boldsymbol{\Phi}_e \;
\text{satisfies RIP} \} \geq 1 - 2 ( n_{p} - 1 )
e^{-C_{4}^{\prime} K} - 2 e^{-C_{4} K_{n_p}}
\end{align}
where $C_{4}^{\prime} = C_{0} - (C_{3} K_{n_p} / K)$ $\times
\left[ 1 + \left( 1 + \log \left( 12 / \delta_S \right) \right) /
\log \left( N/S \right) \right]$, $C_4$ is given after
\eqref{finprob}, and $C_3$ is small enough to guarantee that $C_4$
and $C_{4}^{\prime}$ are both positive.
\end{thmripprobgeneral}

\begin{proof}
See Appendix~E.
\end{proof}

When splitting the rows of $\boldsymbol{\Phi}_e$ in a number of
sets as described before Lemma~\ref{lem:lemripprobgeneral} it may
happen that the last subset
$\left(\boldsymbol{\Phi}_e\right)_{n_{p}}$ has the smallest size
$K_{n_p}$. As a result, the dominant term in
(\ref{eq:finprobgeneral}) will likely be the term $2 e^{-C_{4}
K_{n_p}}$. Moreover, it may lead to a more stringent sparsity
condition, that is, $S \leq C _{3} K_{n_p} / \log(N/S)$. To
improve the lower bound in (\ref{eq:finprobgeneral}), we can move
some of the rows from $\left(\boldsymbol{\Phi}_e\right)_{n_{p}-1}$
to $\left(\boldsymbol{\Phi}_e\right)_{n_{p}}$ in order to make the
last two partitions of almost the same size. Then the requirement
on the sparsity level will become $S \leq C_{3} K^{\prime} /
\log(N/S)$ where  $K^{\prime} = \lfloor (K + K_{n_p} ) / 2
\rfloor$. Therefore, the lower bound on the probability calculated
in (\ref{eq:finprobgeneral}) improves.

\section{Performance Analysis of the recovery}
\label{sec:perfanalysis}
In this section, we aim at answering the question whether signal
recovery also improves if the proposed segmented AIC method, i.e.,
the extended measurement matrix $\boldsymbol{\Phi}_e$
\eqref{eq:largestextendedmeasurementmatrix}, is used instead of
the original matrix $\boldsymbol{\Phi}$. The study is performed
based on the empirical risk minimization method for signal
recovery from noisy random projections \cite{HauptNowak06}. As
mentioned in Section~II, the LASSO method can be viewed as one of
the possible implementations of the empirical risk minimization
method.

We first consider the special case of Example~3 when the extended
measurement matrix is given by (\ref{eq:newmeasurementmatrix}).
Let the entries of the measurement matrix $\boldsymbol{\Phi}$ be
selected with equal probability as $\pm 1/ \sqrt{N}$, i.e., be
i.i.d. Bernoulli distributed with variance $1/N$. This assumption
is the same as in \cite{HauptNowak06} and it is used here in order
to shorten our derivations by only emphasizing the differences
caused by our construction of matrix $\boldsymbol{\Phi}_e$, where
some rows are correlated to each other, as compared to the case
analyzed in \cite{HauptNowak06}, where the measurement matrix
consists of all i.i.d. entries. Note that our results can be
easily applied to the case of Gaussian distributed entries of
$\boldsymbol{\Phi}$ by only changing the moments of Bernoulli
distribution to the moments of Gaussian distribution.

Let $r (\hat{\boldsymbol{f}}, \boldsymbol{f}) \triangleq r (
\hat{\boldsymbol{f}}) - r(\boldsymbol{f} )$ be the ``excess risk''
between the candidate reconstruction $\hat{\boldsymbol{f}}$ of the
signal sampled using the extended measurement matrix
$\boldsymbol{\Phi}_e$ and the actual signal $\boldsymbol{f}$, and
$\hat{r} (\hat{ \boldsymbol{f}}, \boldsymbol{f}) \triangleq
\hat{r}( \hat{\boldsymbol{f}}) - \hat{r} (\boldsymbol{f})$ be the
``empirical excess risk'' between the candidate signal
reconstruction and the actual signal, where $r
(\hat{\boldsymbol{f}})$ and $\hat{r} (\hat{\boldsymbol{f}})$ are
defined in (\ref{eq:riskdefs}). Then the difference between the
``excess risk'' and the ``empirical excess risk'' can be found as
\begin{equation}
r ( \hat{\boldsymbol{f}}, \boldsymbol{f}) - \hat{r} (
\hat{\boldsymbol{f}}, \boldsymbol{f}) = \frac{1}{K_e}
\sum_{j=1}^{K_e} {(U_{j}  -E[U_{j}])}
\label{eq:excessminusempexcess}
\end{equation}
where $U_{j} \triangleq (y_{j} - \boldsymbol{\phi}_{j} \boldsymbol{f})^{2} -
( y_{j} - \boldsymbol{\phi}_{j} \hat{\boldsymbol{f}})^{2}$.

The mean-square error (MSE) between the candidate reconstruction
and the actual signal can be expressed as \cite{Vapnik98}
\begin{equation}
{\rm MSE} \triangleq E \left\{ \| \boldsymbol{g} \|^2
\right\} = N r ( \hat{\boldsymbol{f}}, \boldsymbol{f})
\end{equation}
where $\boldsymbol{g} \triangleq \hat{\boldsymbol{f}} -
\boldsymbol{f}$. Therefore, if we know an upper bound on the
right-hand side of \eqref{eq:excessminusempexcess}, denoted
hereafter as $U$, we can immediately find an upper bound on the
MSE in the form ${\rm MSE} \leq N\hat{r} ( \hat{\boldsymbol{f}},
\boldsymbol{f}) + NU$. In other words, to find the candidate
reconstruction $\hat{\boldsymbol{f}}$ one can minimize $\hat{r} (
\hat{\boldsymbol{f}}, \boldsymbol{f}) + U$, that will also result
in a bound on the MSE as in \eqref{eq:empriskminerror}.

The Craig-Bernstein inequality \cite{HauptNowak06}, \cite{Craig33}
can be used in order to find an upper bound $U$ on the right-hand
side of \eqref{eq:excessminusempexcess}. In the notations of our
paper, this inequality states that the probability of the
following event
\begin{equation}
\frac{1}{K_e} \sum_{j=1}^{K_e} {(U_{j}  - E \{ U_{j}\} )} \leq
\frac{\log \left(\frac{1}{\delta} \right)}{K_{e} \epsilon} +
\frac{\epsilon \, \text{var} \left\{ \sum_{j=1}^{K_e} {U_{j}}
\right\} }{2 K_e (1 - \zeta)} \label{eq:craigbernsteinineq}
\end{equation}
is greater than or equal to $1-\delta$ for $0 < \epsilon h \leq
\zeta < 1$, if the random variables $U_j$ satisfy the
following moment condition for some $h > 0$ and all $k \geq 2$
\begin{equation}
E \left\{ \left| U_{j} - E \{ U_j \} \right|^k \right\} \leq
\frac{k! \, \text{var} \{ U_j \} \, h^{k-2} }{2}.
\label{eq:craigbernsteinmomentcond}
\end{equation}
The second term in the right-hand side of
\eqref{eq:craigbernsteinineq} contains the variance $\text{var}
\left\{ \sum_{j=1}^{K_e} {U_{j}}\right\}$, which we need to
calculate or at least find an upper bound on it.

In the case of the extended measurement matrix, the random
variables $U_j$, $j = 1, \hdots, K_e$ all satisfy the moment
condition for the Craig-Bernstein inequality \cite{Craig33} with
the same coefficient $h = 16 B^{2}e + 8 \sqrt{2} B \sigma$, where
$\sigma^2$ is the variance of the Gaussian noise.\footnote{The
derivation of the coefficient $h$ coincides with a similar
derivation in \cite{HauptNowak06}, and therefore, is omitted.}
Moreover, it is easy to show that the following bound on the
variance of $U_j$ is valid for the extended measurement
matrix\footnote{This bound also coincides with a similar one in
\cite{HauptNowak06}}
\begin{equation}
\text{var} \{ U_j \} \leq \left( 2 \frac{\| \boldsymbol{g}
\|^2}{N} + 4 \sigma^2 \right) \frac{\| \boldsymbol{g} \|^2}{N}
\leq \left( 8 B^{2} + 4 \sigma^2 \right) r(
\hat{\boldsymbol{f}}, \boldsymbol{f}) . \label{eq:variancebound}
\end{equation}

However, unlike \cite{HauptNowak06}, in the case of the extended
measurement matrix, the variables $U_j$ are not independent from
each other. Thus, we can not simply replace the term $\text{var}
\left\{ \sum_{j=1}^{K_e} {U_{j}} \right\}$ with the sum of the
variances for $U_{j}, \; j=1, \hdots, K_e$. Using the definition
of the variance, we can write that
\begin{eqnarray}
&& \text{var} \left\{ \sum_{j=1}^{K_{e}} {U_{j}} \right\}
\triangleq E \left\{ \left( \sum_{j=1}^{K_{e}} {U_{j}} \right)^2
\right\} - \left( E \left\{ \sum_{j=1}^{K_{e}}
{U_{j}} \right\} \right)^2 \nonumber \\
&& \quad = \sum_{j=1}^{K_{e}} {E \{ U_{j}^2 \} } + 2
\sum_{i=1}^{K_{e}-1}\sum_{j=i+1}^{K_{e}} {E \{ U_{i} U_{j} \} } -
K_{e}^{2} \left( \frac{\|\boldsymbol{g}\|^2}{N} \right)^2
\nonumber \\
&& \quad = \sum_{j=1}^{K_{e}} {\left( E \{ U_{j}^2 \} - \left(
\frac{\|\boldsymbol{g}\|^2}{N} \right)^2 \right)} + 2
\sum_{i=1}^{K_{e}-1}\sum_{j=i+1}^{K_{e}} {\left( E \{ U_{i}U_{j}
\} - \left( \frac{\|\boldsymbol{g}\|^2}{N} \right)^2
\right)} \nonumber \\
&& \quad = \sum_{j=1}^{K_{e}} {\text{var} \{ U_{j}\} } + 2
\sum_{i=1}^{K_{e}-1}\sum_{j=i+1}^{K_{e}} {\left( E \{ U_{i}U_{j}
\} - \left( \frac{\|\boldsymbol{g}\|^2}{N} \right)^2 \right)}
\label{eq:variancesigmauiexpansion}
\end{eqnarray}
where the upper bound on ${\text{var} \{ U_{j}\} }$ is given by
\eqref{eq:variancebound}. Using the fact that the random noise
components $w_i$ and $w_j$ are independent from
$\boldsymbol{\phi}_{i} \boldsymbol{g}$ and $\boldsymbol{\phi}_{j}
\boldsymbol{g}$ (see the noisy model \eqref{noisymodel}),
respectively, $E \{ U_{i}U_{j} \}$ can be expressed as
\begin{align}
E \{ U_{i}U_{j} \} \!&=\! E \bigl\{ [ 2 w_{i}
\boldsymbol{\phi}_{i} \boldsymbol{g} \!-\! ( \boldsymbol{\phi}_{i}
\boldsymbol{g})^2 ] [ 2 w_{j} \boldsymbol{\phi}_{j} \boldsymbol{g}
\!-\! (\boldsymbol{\phi}_{j} \boldsymbol{g})^2 ] \bigr\} \nonumber \\
&= 4 E \bigl\{ w_{i} w_{j} \bigr \} E \bigl\{
\boldsymbol{\phi}_{i} \boldsymbol{g} \boldsymbol{\phi}_{j}
\boldsymbol{g} \bigr\} - 2 E \bigl\{ w_{i} \bigr\} E \bigl\{
\boldsymbol{\phi}_{i} \boldsymbol{g} ( \boldsymbol{\phi}_{j}
\boldsymbol{g})^2 \bigr\}
\nonumber \\
& - 2 E \bigl\{ w_{j} \bigr\} E \bigl\{ \boldsymbol{\phi}_{j}
\boldsymbol{g} ( \boldsymbol{\phi}_{i} \boldsymbol{g} )^2 \bigr\}
+ E \bigl\{ (\boldsymbol{\phi}_{i} \boldsymbol{g})^2
(\boldsymbol{\phi}_{j} \boldsymbol{g} )^2 \bigr\}.
\label{eq:uiujexpectationexpansion1}
\end{align}
The latter expression can be further simplified using the fact
that $E \{ w_{i} \} = E \{ w_{j} \} = 0$. Thus, we obtain that
\begin{equation}
E \{ U_{i}U_{j} \} = 4 E \bigl\{ w_{i} w_{j} \bigr\} E \bigl\{
(\boldsymbol{\phi}_{i} \boldsymbol{g} ) ( \boldsymbol{\phi}_{j}
\boldsymbol{g}) \bigr\} + E \bigl\{
(\boldsymbol{\phi}_{i}\boldsymbol{g})^2 (\boldsymbol{\phi}_{j}
\boldsymbol{g} )^2 \bigr\} . \label{eq:uiujexpectationexpansion}
\end{equation}

It is easy to verify that if $\boldsymbol{\phi}_i$ and
$\boldsymbol{\phi}_j$ are independent, 
then $E (U_{i}U_{j}) = E \left\{ (\boldsymbol{\phi}_{i}
\boldsymbol{g})^2 \right\} E \left\{ (\boldsymbol{\phi}_{j}
\boldsymbol{g})^2 \right\}$ $= \left( \|\boldsymbol{g}\|^2 / \ N
\right)^2$ which indeed coincides with \cite{HauptNowak06}.
However, in our case, $\boldsymbol{\phi}_i$ and
$\boldsymbol{\phi}_j$ may depend on each other. If they indeed
depend on each other, they have $L = N/M$ common entries, while
the rest of the entries are independent. In addition, the additive
noise terms $w_i$ and $w_j$ are no longer independent random
variables as well and, thus, $E \bigl\{ w_{i} w_{j} \bigr\} =
\sigma^{2} / M$. Without loss of generality, let the first $L$
entries of $\boldsymbol{\phi}_i$ and $\boldsymbol{\phi}_j$ be the
same, that is,
\begin{align}
\boldsymbol{\phi}_{i} \boldsymbol{g} = \overbrace{g_{1} a_{1} +
\hdots + g_{L} a_{L} }^{A} + \overbrace{ g_{L+1} \phi_{i,L+1} +
\hdots + g_{N} \phi_{i,N}}^{P_i} \\
\boldsymbol{\phi}_{j} \boldsymbol{g} = \overbrace{g_{1} a_{1} +
\hdots + g_{L} a_{L}}^{A} + \overbrace{ g_{L+1} \phi_{j,L+1} +
\hdots + g_{N} \phi_{j,N}}^{P_j}
\end{align}
with ${a_{1},...,a_{L}}$ being the common part between
$\boldsymbol{\phi}_{i}$ and $\boldsymbol{\phi}_{j}$.

Let $\boldsymbol{g}_A$ be a sub-vector of $\boldsymbol{g}$
containing the $L$ elements of $\boldsymbol{g}$ corresponding to
the common part between $\boldsymbol{\phi}_{i}$ and
$\boldsymbol{\phi}_{j}$, and $\boldsymbol{g}_{A^\prime}$ be the
sub-vector comprising the rest of the elements. Then using the
fact that $A$, $P_i$, and $P_j$ are all zero mean independent
random variables, we can express $E \{ (\boldsymbol{\phi}_{i}
\boldsymbol{g} ) (\boldsymbol{\phi}_{j} \boldsymbol{g} ) \}$ from
the first term on the right-hand side of
\eqref{eq:uiujexpectationexpansion} as
\begin{align}
E \{ (\boldsymbol{\phi}_{i} \boldsymbol{g} )
(\boldsymbol{\phi}_{j} \boldsymbol{g} ) \} &= E \{ (A + P_{i} ) (A
+ P_{j}) \} =E \{ A^2 \} + E \{A P_{i} \} + E \{ A P_{j} \} + E \{
P_{i} P_{j} \}
\nonumber \\
&= E \{ A^2 \} = \frac{\left( \sum_{k=1}^{L} {g_{k}^2}
\right)^2}{N} = \frac{ \| \boldsymbol{g}_A \|^2}{N}.
\label{eq:fiigfijgexpectation}
\end{align}
Similar, the second term on the right-hand side of
\eqref{eq:uiujexpectationexpansion} can be expressed as
\begin{equation}
\label{eq:fiigsquaredfijgsquaredexpectation1} E \bigl\{
(\boldsymbol{\phi}_{i} \boldsymbol{g})^2 ( \boldsymbol{\phi}_{j}
\boldsymbol{g})^2 \bigr\} = E \left\{ (A^{2}+P_{i}^2+2AP_{i})
(A^{2}+P_{j}^2+2AP_{j})\right\} .
\end{equation}
Using the facts that $4 E
\bigl\{ w_{i} w_{j} \bigr\} = 4 \sigma^{2} / M$, $E \{ A^2 \} = \|
\boldsymbol{g}_A \|^2 / N$, and $E \{ P_i^2 \} = \|
\boldsymbol{g}_{A^\prime} \|^2 / N$, the expression
\eqref{eq:fiigsquaredfijgsquaredexpectation1} can be further rewritten as
\begin{eqnarray}
E \bigl\{ (\boldsymbol{\phi}_{i} \boldsymbol{g})^2 (
\boldsymbol{\phi}_{j} \boldsymbol{g})^2 \bigr\} \!\!\!&=&\!\!\! E
\left\{ A^{4} + A^{2} P_{i}^{2} + A^{2} P_{j}^{2} + P_{i}^{2}
P_{j}^{2} \right\} = E \{ A^{4} \} + 2 \frac{\| \boldsymbol{g}_A
\|^2}{N} \!\cdot\! \frac{\| \boldsymbol{g}_{A^\prime} \|^2}{N} +
\left( \frac{\|
\boldsymbol{g}_{A^\prime} \|^2}{N} \right)^{2} \nonumber \\
\!\!\!&=&\!\!\! E \{ A^{4} \} + \left( \frac{\| \boldsymbol{g}
\|^2}{N} \right)^2 - \left( \frac{\|\boldsymbol{g}_A \|^2}{N}
\right)^2. \label{eq:fiigsquaredfijgsquaredexpectation}
\end{eqnarray}

Substituting (\ref{eq:fiigfijgexpectation}) and
(\ref{eq:fiigsquaredfijgsquaredexpectation}) into
(\ref{eq:uiujexpectationexpansion}), we obtain that
\begin{equation} \label{newequ}
E \{ U_{i} U_{j} \} = \frac{4\sigma^{2}}{M} \cdot \frac{\|
\boldsymbol{g}_A \|^2}{N} + E \{ A^{4} \} + \left( \frac{\|
\boldsymbol{g} \|^2}{N} \right)^2 - \left( \frac{\|
\boldsymbol{g}_A \|^2}{N} \right)^2.
\end{equation}
Moreover, substituting \eqref{newequ} into
(\ref{eq:variancesigmauiexpansion}), we find that
\begin{equation} \label{eq:varianceofsummationbound1}
\text{var} \left\{ \sum_{j=1}^{K_{e}} {U_{j}} \right\} =
\sum_{j=1}^{K_{e}} {\text{var} \{ U_{j} \} } + 2
\sum_{\boldsymbol{\phi}_{i}, \boldsymbol{\phi}_j \text{dependent}}
{\left( E \{ A^4 \} - \left( \frac{\| \boldsymbol{g}_A
\|^2}{N}\right)^{2} + \frac{4 \sigma^{2}}{M} \cdot \frac{\|
\boldsymbol{g}_A \|^2}{N} \right)}.
\end{equation}

Using the fact that the extended measurement matrix is constructed such that the
waveforms $\boldsymbol{\phi}_{i}$, $i = K+1, \hdots, K_{e}$ are
built upon $M$ rows of the original matrix and also using the
inequality\footnote{We skip the derivation of this inequality since
it is relatively well known and can be found, for example, in
\cite[p.~4039]{HauptNowak06}.} $E \{ A^4 \} - \left( \|
\boldsymbol{g}_A \|^2 / N \right)^2 \leq 2 \left( \|
\boldsymbol{g}_A \|^2 / N \right)^2$ for all these $M$ rows, we
obtain for every $\boldsymbol{\phi}_{i}$, $i = K+1, \hdots, K_{e}$
that
\begin{equation} \label{eqkj}
\sum_{k=1}^{M} {\left( E \{ A^4 \} - \left( \frac{\|
\boldsymbol{g}_A \|^2}{N} \right)^{2} + \frac{4 \sigma^{2}}{M}
\cdot \frac{\| \boldsymbol{g}_A \|^2}{N} \right) } \leq
\sum_{k=1}^{M} {\left( 2 \left( \frac{\| \boldsymbol{g}_A \|^2}{N}
\right)^{2} + \frac{4 \sigma^{2}}{M} \cdot \frac{\|
\boldsymbol{g}_A \|^2}{N} \right) }
\end{equation}
where $\boldsymbol{g}_A$ corresponds to the first $L$ entries of
$\boldsymbol{g}$ for $k=1$, to the entries from $L+1$ to $2L$ for
$k=2$ and so on. Applying also the triangle inequality, we find
that
\begin{equation} \label{eqkj1}
\sum_{k=1}^{M} {\left( 2 \left( \frac{\| \boldsymbol{g}_A \|^2}{N}
\right)^{2} + \frac{4 \sigma^{2}}{M} \cdot \frac{\|
\boldsymbol{g}_A \|^2}{N} \right) } \leq 2 \left( \frac{\|
\boldsymbol{g} \|^2}{N} \right)^{2} + \frac{4\sigma^{2}}{M} \cdot
\frac{\| \boldsymbol{g} \|^2}{N}.
\end{equation}

Combining \eqref{eqkj} and
\eqref{eqkj1} and using the fact that there are $K_a$ additional
rows in the extended measurement matrix, we obtain that
\begin{equation} \label{jgdag}
2 \sum_{\boldsymbol{\phi}_{i}, \boldsymbol{\phi}_j
\text{dependent}} {\left( E \{ A^4 \} - \left( \frac{\|
\boldsymbol{g}_A \|^2}{N}\right)^{2} + \frac{4 \sigma^{2}}{M}
\cdot \frac{\| \boldsymbol{g}_A \|^2}{N} \right)} \leq 4 K_{a}
\left( \frac{\| \boldsymbol{g} \|^2}{N} \right)^{2} + \frac{8
\sigma^{2}K_{a}}{M} \cdot \frac{\| \boldsymbol{g} \|^2}{N}.
\end{equation}
Noticing that $\| \boldsymbol{g} \|^2 / N = r(
\hat{\boldsymbol{f}}, \boldsymbol{f})$ and $\| \boldsymbol{g}
\|^{2} \leq 4 N B^2$, the right-hand side of the inequality
\eqref{jgdag} can be further upper bounded as
\begin{equation} \label{jgdag1}
4 K_{a} \left( \frac{\| \boldsymbol{g} \|^2}{N} \right)^{2} +
\frac{8 \sigma^{2}K_{a}}{M} \cdot \frac{\| \boldsymbol{g} \|^2}{N}
\leq 16 K_{a} B^{2} \, r( \hat{\boldsymbol{f}}, \boldsymbol{f}) +
\frac{8 \sigma^{2} K_{a}}{M} \, r(\hat{\boldsymbol{f}},
\boldsymbol{f}) .
\end{equation}

Using the upper bound \eqref{jgdag1} for the second term in
(\ref{eq:varianceofsummationbound1}) and the upper bound
(\ref{eq:variancebound}) for the first term in
(\ref{eq:varianceofsummationbound1}), we finally can upper bound
the $\text{var} \left\{ \sum_{j=1}^{K_{e}} {U_{j}} \right\}$ as
\begin{equation} \label{eq:varianceofsummationbound}
\text{var} \left\{ \sum_{j=1}^{K_{e}} {U_{j}} \right\} \leq K_{e}\left(
8 B^{2} \left( 1 + \frac{2 K_{a}} {K_{e}} \right) + 4 \sigma^{2}
\left( 1 + \frac{2K_{a}}{MK_{e}} \right) \right) \,
r(\hat{\boldsymbol{f}}, \boldsymbol{f}).
\end{equation}
Therefore, based on the Craig-Bernstein inequality, the
probability that for a given candidate signal $\hat{\boldsymbol{f}}$
the following inequality holds
\begin{equation} \label{CBfinalpre}
r ( \hat{\boldsymbol{f}}, \boldsymbol{f}) - \hat{r} (
\hat{\boldsymbol{f}}, \boldsymbol{f}) \leq \frac{\log(
\frac{1}{\delta} )}{K_{e}\epsilon} + \frac{\left( 8 B^{2} \left( 1
+ \frac{2 K_{a}}{K_e} \right) + 4 \sigma^{2} \left( 1 + \frac{2
K_{a}}{M K_{e}} \right) \right) r(\hat{\boldsymbol{f}},
\boldsymbol{f}) \, \epsilon}{2 ( 1 - \zeta )}
\end{equation}
is greater than or equal to $1-\delta$.

Let $c ( \hat{\boldsymbol{f}} )$ be chosen such that the Kraft
inequality $\sum_{\hat{\boldsymbol{f}} \in {\cal F} (B)}{2^{c (
\hat{\boldsymbol{f}} )}} \leq 1$ is satisfied (see also
\cite{HauptNowak06}), and let $\delta ( \hat{\boldsymbol{f}}) =
2^{-c (\hat{\boldsymbol{f}} )} \, \delta$. Applying the union
bound to \eqref{CBfinalpre}, it can be shown that for all
$\hat{\boldsymbol{f}} \in {\cal F} (B)$ and for all $\delta > 0$,
the following inequality holds with probability of at least
$1-\delta$
\begin{align}
r ( \hat{\boldsymbol{f}}, \boldsymbol{f} ) - \hat{r} (
\hat{\boldsymbol{f}}, \boldsymbol{f}) \leq \frac{c
(\hat{\boldsymbol{f}} ) \log 2 + \log ( \frac{1}{\delta} )}{K_{e}
\, \epsilon} + \frac{\left( 8 B^{2} \left( 1 + \frac{2
K_{a}}{K_{e}} \right) + 4 \sigma^{2} \left( 1 + \frac{2
K_{a}}{MK_{e}} \right) \right) \, r ( \hat{\boldsymbol{f}},
\boldsymbol{f}) \, \epsilon}{2 ( 1 - \zeta)} .
\end{align}
Finally, setting $\zeta =
\epsilon \, h$ and
\begin{align} \label{coefA}
&a = \frac{\left( 8 B^{2} \left( 1 + \frac{2 K_{a}}{K_e} \right)
+ 4 \sigma^{2} \left( 1 + \frac{2 K_{a}}{MK_{e}} \right) \right)
\, \epsilon}{2 ( 1 - \zeta)} \\
&\epsilon < \frac{1}{\left( 4 \left( 1 + \frac{2K_{a}}{K_e} \right) +
16 e \right) B^{2} + 8 \sqrt{B} \sigma + 2 \sigma^2 \left( 1 +
\frac{2 K_{a}}{MK_{e}} \right)} \label{coefEpsilon}
\end{align}
where $0 < \epsilon \, h \leq \zeta < 1$ as required by the
Craig-Bernstein inequality, the following inequality holds with
probability of at least $1 - \delta$ for all $\hat{\boldsymbol{f}}
\in {\cal F} (B)$
\begin{align}
\label{eq:boundonr}
( 1 - a ) r ( \hat{\boldsymbol{f}}, \boldsymbol{f}) \leq \hat{r}(
\hat{\boldsymbol{f}}, \boldsymbol{f}) + \frac{c (
\hat{\boldsymbol{f}} ) \log 2 + \log ( \frac{1}{\delta} )}{K_{e}
\, \epsilon}.
\end{align}

The following result on the recovery performance of the empirical risk
minimization method is in order.

\newtheorem{thmriskminerrbnd}[kminus1newsets]{Theorem}
\begin{thmriskminerrbnd}
\label{thm:thmriskminerrbnd} Let $\epsilon$ be chosen as
\begin{equation}
    \epsilon = \frac{1}{\left( 60 \left(B + \sigma \right)^2 \right)}
    \label{eq:epsilonforke}
\end{equation}
which satisfies the inequality
\eqref{coefEpsilon}, then the signal reconstruction
$\hat{\boldsymbol{f}}_{K_e}$ given by
\begin{equation}
\hat{\boldsymbol{f}}_{K_e} = \arg \underset{ \hat{\boldsymbol{f}}
\in {\cal F} (B)} {\min} \left\{ \hat{r} ( \hat{\boldsymbol{f}} ) +
\frac{c (\hat{\boldsymbol{f}}) \log 2}{\epsilon K_{e}} \right\}
\label{eq:emprisminkextended}
\end{equation}
satisfies the following inequality
\begin{equation}
E \left\{ \frac{\| \hat{\boldsymbol{f}}_{K_e} - \boldsymbol{f}
\|^2}{N} \right\} \leq C_{1e} \underset {\hat{\boldsymbol{f}} \in
{\cal F} (B)}{\min} \left\{ \frac{\| \hat{\boldsymbol{f}} -
\boldsymbol{f} \|^2}{N} + \frac{c (\hat{\boldsymbol{f}}) \log
2 + 4}{\epsilon K_{e}} \right\}
\end{equation}
where $C_{1e}$ is the constant given as
\begin{equation} \label{C1e}
C_{1e} = \frac{1+a}{1-a}, \quad a = \frac{2 \left( 1 +
\frac{2K_{a}} {K_{e}} \right) \left( \frac{B}{\sigma} \right)^{2}
+ \left( 1 + \frac{2K_{a}}{MK_{e}} \right)} {\left( 30 - 8 e
\right) \left( \frac{B}{\sigma} \right)^{2} + \left( 60 - 4
\sqrt{2} \right) \left( \frac{B} {\sigma} \right) + 30}
\end{equation}
with $a$ obtained from \eqref{coefA} for the specific choice of
$\epsilon$ in \eqref{eq:epsilonforke}.
\end{thmriskminerrbnd}

\begin{proof}
The proof follows the exact steps of the proof of the related
result for the uncorrelated case
\cite[p.~4039--4040]{HauptNowak06} with the exception of using, in
our correlated case, the above calculated values for  $\epsilon$
\eqref{eq:epsilonforke} and $a$ \eqref{C1e}.
\end{proof}

{\bf Example 4:} Let one set of samples be obtained based on the
measurement matrix $\boldsymbol{\Phi}_e$ with $K_{a} = K$, $K_e =
2K$, and $M = 8$, and let another set of samples be obtained using
a $2K\times N$ measurement matrix with all i.i.d. (Bernoulli)
elements. Let also $\epsilon$ be selected as
\eqref{eq:epsilonforke}. Then the MSE error bounds for these two
cases differ from each other only by a constant factor given for
the former case by $C_{1e}$ in \eqref{C1e} and in the latter case
by $C_1$ (see \eqref{eq:empriskminerror} and the row after).
Considering the two limiting cases when $B / \sigma \rightarrow 0$
and $B / \sigma \rightarrow \infty$, the intervals of change for
the corresponding coefficients can be obtained as $1.08 \leq
C_{1e} \leq 2.88$ and $1.06 \leq C_{1} \leq 1.63$, respectively.

The following result on the achievable recovery performance for a
sparse or compressible signal sampled based on the extended
measurement matrix $\boldsymbol{\Phi}_e$ is also of great
interest.

\newtheorem{thmriskminerrbndsprs}[kminus1newsets]{Theorem}
\begin{thmriskminerrbndsprs}
\label{thm:thmriskminerrbndsprs} For a sparse signal
${\boldsymbol{f} \in {\cal F}_{s}(B,S)} = \{ \boldsymbol{f}: \|
\boldsymbol{f} \|^{2} \leq N B^{2}, \| \boldsymbol{f}\|_{l_0} \leq
S \}$ and corresponding reconstructed signal
$\hat{\boldsymbol{f}}_{K_e}$ obtained according to
\eqref{eq:emprisminkextended}, there exists a constant
$C^{\prime}_{2e} = C^{\prime}_{2e} (B,\sigma) > 0$, such that
\begin{equation}
\underset{ \boldsymbol{f} \in {\cal F}_{s} (B,S)} { \sup} E
\left\{ \frac{\| \hat{\boldsymbol{f}}_{K_e} - \boldsymbol{f}
\|^2}{N} \right\} \leq C_{1e} C^{\prime}_{2e} \left( \frac{K_e}{S
\log N} \right)^{-1} . \label{eq:newcomperror}
\end{equation}

Similar, for a compressible signal $\boldsymbol{f} \in { \cal
F}_{c} ( B, \alpha, C_{A} ) = \{ \boldsymbol{f}: \| \boldsymbol{f}
\|^2 \leq N B^{2}, \| \boldsymbol{f}^{(m)} - \boldsymbol{f} \|^2$
$\leq N C_{A} m^{-2\alpha} \}$ and corresponding reconstructed
signal $\hat{\boldsymbol{f}}_{K_e}$ obtained according to
\eqref{eq:emprisminkextended}, there exists a constant $C_{2e} =
C_{2e} (B,\sigma,C_{A}) > 0$, such that
\begin{equation}
\underset{ \boldsymbol{f} \in {\cal F}_{c} (B,\alpha,C_{A})}
{\sup} E \left\{ \frac{\| \hat{\boldsymbol{f}}_{K_e} - \boldsymbol{f}
\|^2}{N} \right\} \leq C_{1e} C_{2e}
\left( \frac{K_e}{\log N} \right)^{-2 \alpha/(2 \alpha + 1)} .
\label{eq:supcomperrorextended}
\end{equation}
\end{thmriskminerrbndsprs}

\begin{proof}
The proof follows the exact steps of the proofs of the related
results for the uncorrelated case
\cite[p.~4040--4041]{HauptNowak06} with the exception of using, in
our correlated case, the above calculated values for $\epsilon$
\eqref{eq:epsilonforke} and $a$ \eqref{C1e}.
\end{proof}

\begin{figure}[t]
\centering
\includegraphics[scale=0.6]{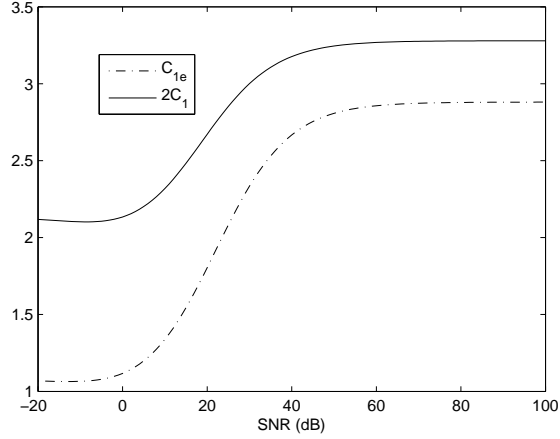}
\caption{$C_{1e}$ and $2C_{1}$ versus SNR.}
\label{fig:c1evs2c1}
\end{figure}

{\bf Example 5:} Let one set of samples be obtained based on the
extended measurement matrix $\boldsymbol{\Phi}_e$ with $K_{a} =
K$, $K_e = 2K$, and $M = 8$ and let another set of samples be
obtained using the $K \times N$ measurement matrix with all i.i.d.
(Bernoulli) elements. The error bounds corresponding to the case
of $K$ uncorrelated samples of \cite{HauptNowak06} and our case of
$K_e$ correlated samples are (\ref{eq:originalcomperror}) and
\eqref{eq:newcomperror}, respectively. The comparison between
these two error bounds boils down in this example to comparing
$2C_{1} C^{\prime}_2$ and $C_{1e} C^{\prime}_{2e}$. Assuming the
same $\epsilon$ as \eqref{eq:epsilonforke} for both methods, the
following holds true $C^{\prime}_{2e}=C^{\prime}_2$.
Fig.~\ref{fig:c1evs2c1} compares $C_{1e}$ and $2C_{1}$ versus the
signal-to-noise ratio (SNR) $B^{2} / \sigma^{2}$. Since $C_{1e} <
2C_{1}$ for all values of SNR, the quality of the signal recovery,
i.e., the corresponding MSE, for the case of $2K \times N$
extended measurement matrix is expected to be better than the
quality of the signal recovery for the case of $K \times N$
measurement matrix of all i.i.d. entries.

The above results can be easily generalized for the case when $K_{a}
> K$. Indeed, we only need to recalculate $\text{var} \left\{
\sum_{j=1}^{K_{e}} {U_{j}} \right\}$ for $K_{a} > 2K$. The only
difference with the previous case of $K_{a} \leq K$ is the
increased number of pairs of dependent rows in the extended
measurement matrix $\boldsymbol{\Phi}_e$, which has a larger size
now. The latter affects only the second term in
\eqref{eq:varianceofsummationbound1}. In particular, every row in
$\boldsymbol{\Phi}^{{\cal P}^{(1)}}$ depends on $M$ rows of the
original measurement matrix $\boldsymbol{\Phi}$. Moreover, the
term $\sum_{i=1}^{2K-1}\sum_{j=i+1}^{2K} {E \{ U_{i} U_{j} \} }$
over all these $M$ rows is bounded as in \eqref{eqkj1}. Then
considering all $K M$ pairs of dependent rows from
$\boldsymbol{\Phi}$ and $\boldsymbol{\Phi}^{{\cal P}^{(1)}}$, we
have
\begin{equation} \label{gen1}
2 \sum_{\boldsymbol{\phi}_{i}, \boldsymbol{\phi}_j
\text{dependent}} {\left( E \{ A^4 \} - \left( \frac{\|
\boldsymbol{g}_A \|^2}{N}\right)^{2} + \frac{4 \sigma^{2}}{M}
\cdot \frac{\| \boldsymbol{g}_A \|^2}{N} \right)} \leq 4 K \left(
\frac{\| \boldsymbol{g} \|^2}{N} \right)^{2} + \frac{8
\sigma^{2}K}{M} \cdot \frac{\| \boldsymbol{g} \|^2}{N}.
\end{equation}

Similar, every row of $\boldsymbol{\Phi}^{{\cal P}^{(2)}}$ depends
on $M$ rows of $\boldsymbol{\Phi}^{{\cal P}^{(1)}}$ and $M$ rows
of $\boldsymbol{\Phi}$. Considering all these $2 K M$ pairs of
dependent rows, we have
\begin{equation} \label{gen2}
2 \sum_{\boldsymbol{\phi}_{i}, \boldsymbol{\phi}_j
\text{dependent}} {\left( E \{ A^4 \} - \left( \frac{\|
\boldsymbol{g}_A \|^2}{N}\right)^{2} + \frac{4 \sigma^{2}}{M}
\cdot \frac{\| \boldsymbol{g}_A \|^2}{N} \right)} \leq 4 (2K)
\left( \frac{\| \boldsymbol{g} \|^2}{N} \right)^{2} + \frac{8
\sigma^{2}(2K)}{M} \cdot \frac{\| \boldsymbol{g} \|^2}{N} .
\end{equation}

Finally, the number of rows in the last matrix
$\left(\boldsymbol{\Phi}_{e}\right)_{n_{p}}$ is $K_{n_p}$ (see
\eqref{eq:numofpartsgeneral} and
\eqref{eq:extendedmeasurementmatrixgeneral}). Every row of
$\left(\boldsymbol{\Phi}_{e}\right)_{n_{p}}$ depends on $M$ rows
of each of the previous $n_{p}-1$ matrices
$\boldsymbol{\Phi}^{{\cal P}^{(i)}}$, $i= 1, \hdots, n_{p}-1$.
Considering all $(n_{p} -1) K_{n_p} M$ pairs of dependent rows, we
have
\begin{equation} \label{gen3}
2 \!\!\! \sum_{\boldsymbol{\phi}_{i}, \boldsymbol{\phi}_j
\text{dependent}} {\left( E \{ A^4 \} \!-\! \left( \frac{\|
\boldsymbol{g}_A \|^2}{N}\right)^{2} \!+\! \frac{4 \sigma^{2}}{M}
\cdot \frac{\| \boldsymbol{g}_A \|^2}{N} \right)} \leq 4 (n_{p}
\!-\! 1)K_{n_p} \left( \frac{\| \boldsymbol{g} \|^2}{N}
\right)^{2} \!+\! \frac{8 \sigma^{2}(n_{p} \!-\! 1)K_{n_p}}{M}
\cdot \frac{\| \boldsymbol{g} \|^2}{N}.
\end{equation}

Based on the equations \eqref{eq:variancesigmauiexpansion} and
\eqref{gen1}--\eqref{gen3} we can find the following bound
\begin{equation} \label{eq:varianceofsummationboundgeneral}
\text{var} \left\{ \sum_{j=1}^{K_{e}} {U_{j}} \right\} \leq
K_{e}\left( 8 B^{2} \left( 1 + \frac{D} {K_{e}} \right) + 4
\sigma^{2} \left( 1 + \frac{D}{MK_{e}} \right) \right) \,
r(\hat{\boldsymbol{f}}, \boldsymbol{f})
\end{equation}
where $D = 2K\sum_{i=1}^{n_{p}-2}{i}+2K_{n_p}(n_{p}-1)$.
Note that in the case that $K_e=n_{p}K$, we have $D/K_e = n_{p}-1$.

Therefore, it can be shown for the general extended matrix
\eqref{eq:largestextendedmeasurementmatrix} that the inequality
\eqref{eq:boundonr} holds with the following values of $a$ and
$\epsilon$:
\begin{align} \label{coefAgeneral}
&a = \frac{\left( 8 B^{2} \left( 1 + \frac{D}{K_e} \right)
+ 4 \sigma^{2} \left( 1 + \frac{D}{MK_{e}} \right) \right)
\, \epsilon}{2 ( 1 - \zeta)} \\
&\epsilon < \frac{1}{\left( 4 \left( 1 + \frac{D}{K_e} \right) +
16 e \right) B^{2} + 8 \sqrt{B} \sigma + 2 \sigma^2 \left( 1 +
\frac{D}{MK_{e}} \right)} \label{coefEpsilongeneral}
\end{align}
Moreover, the theorems similar to
Theorems~\ref{thm:thmriskminerrbnd}~and~\ref{thm:thmriskminerrbndsprs}
follow straightforwardly with the corrections to $a$ and
$\epsilon$ which are given now by \eqref{coefAgeneral} and
\eqref{coefEpsilongeneral}, respectively.

We finally make some remarks on {\it non-RIP} conditions for
$l_1$-norm-based recovery. Since the extended measurement matrix
of the proposed segmented compressed sampling method satisfies the
RIP, the results of \cite{CandesRombergTao05} on recoverability
and stability of the $l_1$-norm minimization straightforwardly
apply. A different non-RIP-based approach for studying the
recoverability and stability of the $l_1$-norm minimization, which
uses some properties of the null space of the measurement matrix,
is used in \cite{YinZhang}. Then the non-RIP sufficient condition
for recoverability of a sparse signal from its noiseless
compressed samples with the algorithm
\eqref{eq:loneminimizationchanged} is \cite{YinZhang}
\begin{equation}
\sqrt{S} < \min \left\{0.5\frac{\|{\boldsymbol v} \|_{l_1}}{\|
{\boldsymbol v}\|_{l_2}}:\; {\boldsymbol v} \in \left\{ {\cal N}
(\boldsymbol{\Phi}) \setminus \{0\} \right\} \right\}
\label{eq:suffrecovcondition}
\end{equation}
where ${\cal N}(\boldsymbol{\Phi})$ denotes the null space of the
measurement matrix $\boldsymbol{\Phi}$.

Let us show that the condition \eqref{eq:suffrecovcondition} is
also satisfied for the extended measurement matrix
$\boldsymbol{\Phi}_e$. Let $\boldsymbol{d}$ be any vector in the
null space of $\boldsymbol{\Phi}_e$, i.e., $\boldsymbol{d} \in
{\cal N} (\boldsymbol{\Phi}_e)$. Therefore,
$[{\boldsymbol{\Phi}_e}]_i \boldsymbol{d} = 0, \; i=1, \hdots,
K_e$ where $[{\boldsymbol{\Phi}_e}]_i$ is the $i$-th $1 \times N$
row-vector of $\boldsymbol{\Phi}_e$. Since the first $K$ rows of
$\boldsymbol{\Phi}_e$ are exactly the same as the $K$ rows of
$\boldsymbol{\Phi}$, we have $[\boldsymbol{\Phi}]_i \boldsymbol{d}
= 0, \; i=1,\hdots,K$. Therefore, $\boldsymbol{d} \in {\cal N}
(\boldsymbol{\Phi})$, and we can conclude that ${\cal N}
(\boldsymbol{\Phi}_e) \subset {\cal N}(\boldsymbol{\Phi})$. Due to
this property, we have $\min \left\{ 0.5 \|  {\boldsymbol v}
\|_{l_1} /\| {\boldsymbol v} \|_{l_2}: \;  {\boldsymbol v} \in
{\cal N}(\boldsymbol{\Phi})\right\} \leq \min \left\{0.5 \|
{\boldsymbol v} \|_{l_1} / \|  {\boldsymbol v} \|_{l_2}: \;
{\boldsymbol v} \in {\cal N} (\boldsymbol{\Phi}_e) \right\}$.
Therefore, if the original measurement matrix $\boldsymbol{\Phi}$
satisfies \eqref{eq:suffrecovcondition}, so does the extended
measurement matrix $\boldsymbol{\Phi}_e$, and the signal is
recoverable from the samples taken by $\boldsymbol{\Phi}_e$.

Moreover, the necessary and sufficient condition for all signals
with $\| \boldsymbol{x} \|_{l_0} < S$ to be recoverable from noiseless
compressed samples using the $l_1$-norm minimization
\eqref{eq:loneminimizationchanged} is that \cite{YinZhang}
\begin{equation}
\| {\boldsymbol v} \|_{l_1} > 2 \| {\boldsymbol v}_{\cal T}
\|_{l_1}, \; \forall {\boldsymbol v} \in \left\{ {\cal N}
(\boldsymbol{\Phi}) \setminus \{0\} \right\}
    \label{eq:necesssuffrecovcondition}
\end{equation}
where $\cal T$ is the set of indexes corresponding to the nonzero
coefficients of $\boldsymbol{x}$. It is easy to see that since
${\cal N}(\boldsymbol{\Phi}_e) \subset {\cal N}
(\boldsymbol{\Phi})$, the condition
\eqref{eq:necesssuffrecovcondition} also holds for the extended
measurement matrix if the original measurement matrix satisfies
it.

\section{Simulation Results}
\label{sec:simresults} Throughout our simulations we use the
sparse signal of dimension 128 with only 3 nonzero entries, which
are set to $\pm 1$ with equal probabilities. Since the signal is
sparse in the time domain, $\boldsymbol{\Psi} = \boldsymbol{I}$.
The collected samples are assumed to be noisy, i.e., the model
\eqref{noisymodel} is applied. In all our simulation examples,
three different measurement matrices (sampling schemes) are used:
(i)~the $K \times N$ measurement matrix $\boldsymbol{\Phi}$ with
i.i.d. entries referred to as the original measurement matrix;
(ii)~the extended $K_{e} \times N$ measurement matrix
$\boldsymbol{\Phi}_e$ obtained using the proposed segmented
compressed sampling method and referred to as the extended
measurement matrix; and (iii)~the $K_{e} \times N$ measurement
matrix with all i.i.d entries referred to as the enlarged
measurement matrix. This last measurement matrix corresponds to
the sampling scheme with $K_{e}$ independent BMIs in the AIC in
Fig.~\ref{fig:parallelaicstructure}. The number of segments in the
proposed segmented compressed sampling method $M$ is set to $8$.
To make sure that the measurement noise for additional samples
obtained based on the extended measurement matrix is correlated
with the measurement noise of the original samples, the $K \times
M$ matrix of noisy sub-samples with the noise variance
$\sigma^{2}/M$ is first generated. Then the permutations are
applied to this matrix and the sub-samples along each row of the
original and permuted matrices are added up together to build the
noisy samples.

The recovery performance for three aforementioned sampling schemes
is measured using the MSE between the
recovered and original signals. In all examples, MSE values are computed
based on 5000 independent simulation runs for all sampling schemes
tested. The SNR is defined as $\| \boldsymbol{\Phi} \boldsymbol{f}
\|_{l_{2}}^{2} / \| \boldsymbol{w} \|_{l_{2}}^{2}$. Approximating
$\|\boldsymbol{\Phi} \boldsymbol{f} \|_{l_{2}}^{2}$ by $(K^\prime
/ N) \| \boldsymbol{f} \|_{l_{2}}^{2}$, which is valid because of
(\ref{eq:ripconstant}), the corresponding noise variance
$\sigma^2$ can be calculated if SNR is given, and vise versa. Here
$K^\prime = K$ for the sampling scheme based on the original
measurement matrix, while $K^\prime = K_e$ in the other two schemes. For
example, the approximate SNR in dBs can be calculated as $10
\log_{10}{( 3 / {N \sigma^2})}$.

{\it Recovery based on the $l_1$-norm minimization algorithm:} In
our first simulation example, the $l_1$-norm minimization
algorithm \eqref{eq:noisyloneminimization} is used to recover a
signal sampled using the three aforementioned sampling schemes.
Since $\boldsymbol{\Psi} = \boldsymbol{I}$, then
$\boldsymbol{\Phi} ^{\prime}= \boldsymbol{\Phi}$ in
\eqref{eq:noisyloneminimization}. The number of BMIs in the
sampling device is taken to be $K = 16$, while $\gamma$ in
\eqref{eq:noisyloneminimization}, which is the bound on the root
square of the noise energy, is set to $\sqrt{K^{\prime}} \sigma$.
The entries of the original and enlarged measurement matrices are
generated as i.i.d. Gaussian distributed random variables with
zero mean and variance $1/N$.

\begin{figure}
\centering
\includegraphics[scale=0.47]{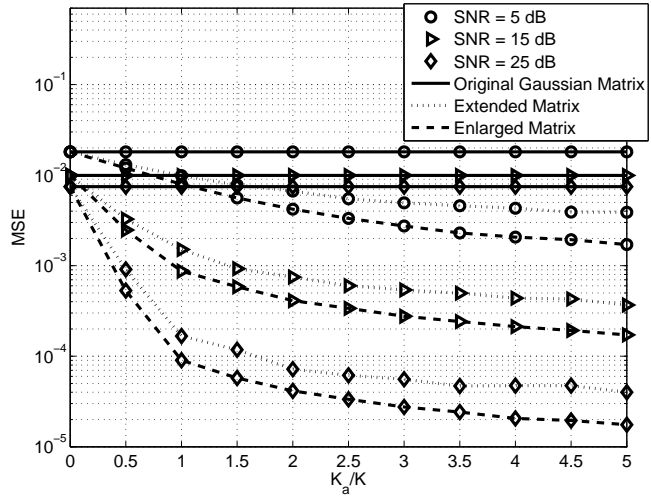}
\caption{Recovery based on the $l_1$-norm minimization algorithm:
MSEs versus $K_{a} / K$.} \label{fig:l1gaussianmsevska}
\end{figure}

Fig.~\ref{fig:l1gaussianmsevska} shows the MSEs corresponding to
all three aforementioned measurement matrices versus the ratio of
the number of additional samples to the number of original samples
$K_{a} / K$. The results are shown for three different SNR values
of 5, 15 and 25~dB. It can be seen from the figure that better
recovery quality is achieved by using the extended measurement
matrix as compared to the original measurement matrix. The
improvements are more significant for high SNRs since the recovery
error is proportional to the noise power
\cite{CandesRombergTao05}. As expected, the recovery performance
in the case of the extended measurement matrix is not as good as
in the case of the enlarged measurement matrix. This difference,
however, is small as compared to the performance improvement over
the original measurement matrix. Note also that in the case of the
enlarged measurement matrix, the AIC in
Fig.~\ref{fig:parallelaicstructure} consists of $K_e$ BMIs, while
only $K$ BMIs are required in the case of the extended measurement
matrix. Thus, the segmented AIC requires $K_e - K$ less BMIs. For
example, the number of such BMIs halves if $K_{a} / K = 1$.
Additionally, it can be seen that the rate of MSE improvement
decreases as the number of collected samples increases. The latter
can be observed for both the extended and enlarged measurement
matrices and for all three values of SNR.

{\it Recovery based on the empirical risk minimization method:} In
our second simulation example, the empirical risk minimization
method is used to recover a signal sampled using the three
aforementioned sampling schemes tested with $K = 24$. The
minimization problem \eqref{eq:empriskminequation} is solved to
obtain a candidate reconstruction
$\hat{\boldsymbol{f}}_{K^\prime}$ of the original sparse signal
$\boldsymbol{f}$. Considering $\hat{\boldsymbol{f}}_{K^\prime} =
\boldsymbol{\Psi}^{H} \hat{\boldsymbol{x}}_{K^\prime}$, the
problem \eqref{eq:empriskminequation} can be rewritten in terms of
$\hat{\boldsymbol{x}}_{K^\prime}$ as
\begin{equation}
\hat{\boldsymbol{x}}_{K^\prime} = \arg \underset{\hat{\boldsymbol{x}} \in
{\cal X}}{\min}\left\{\hat{r}(\boldsymbol{\Psi}^{H}
\hat{\boldsymbol{x}} ) +  \frac{c(\hat{\boldsymbol{x}}) \log
2}{\epsilon {K^\prime}}\right\} = \arg \underset{\hat{\boldsymbol{x}} \in
{\cal X}}{\min}\left\{\| \boldsymbol(y) - \boldsymbol{\Phi}
\boldsymbol{\Psi}^{H}\hat{\boldsymbol{x}}\|_{l_2}^{2} + \frac{2
\log 2 \log N}{\epsilon} \|\hat{\boldsymbol{x}}\|_{l_0}\right\}
\label{eq:empriskminequationinpsibasis}
\end{equation}
and solved using the iterative bound optimization procedure
\cite{HauptNowak06}. This procedure uses the threshold $\sqrt{2
\log 2 \log N / \lambda \epsilon}$, where $\lambda$ is the largest
eigenvalue of the matrix $\boldsymbol{\Phi}^{T}
\boldsymbol{\Phi}$. In our simulations, this threshold is set to
0.035 for the case of the extended measurement matrix and 0.05 for
the cases of the original and enlarged measurement matrices. These
threshold values are optimized as recommended in
\cite{HauptNowak06}. The stopping criterion for the iterative
bound optimization procedure is $\| \hat{\boldsymbol{x}}^{(i+1)} -
\hat{\boldsymbol{x}}^{(i)} \|_{l_{\infty}} \leq \theta$, where
$\|.\|_{l_{\infty}}$ is the $l_{\infty}$ norm and
$\hat{\boldsymbol{x}}^{(i)}$ denotes the value of
$\hat{\boldsymbol{x}}$ obtained in the $i$-th iteration. The value
$\theta = 0.001$ is selected.

\begin{figure}
\centering \subfigure[Measurement matrix with Gaussian distributed entries] {
\includegraphics[angle=0,width=0.47\textwidth]{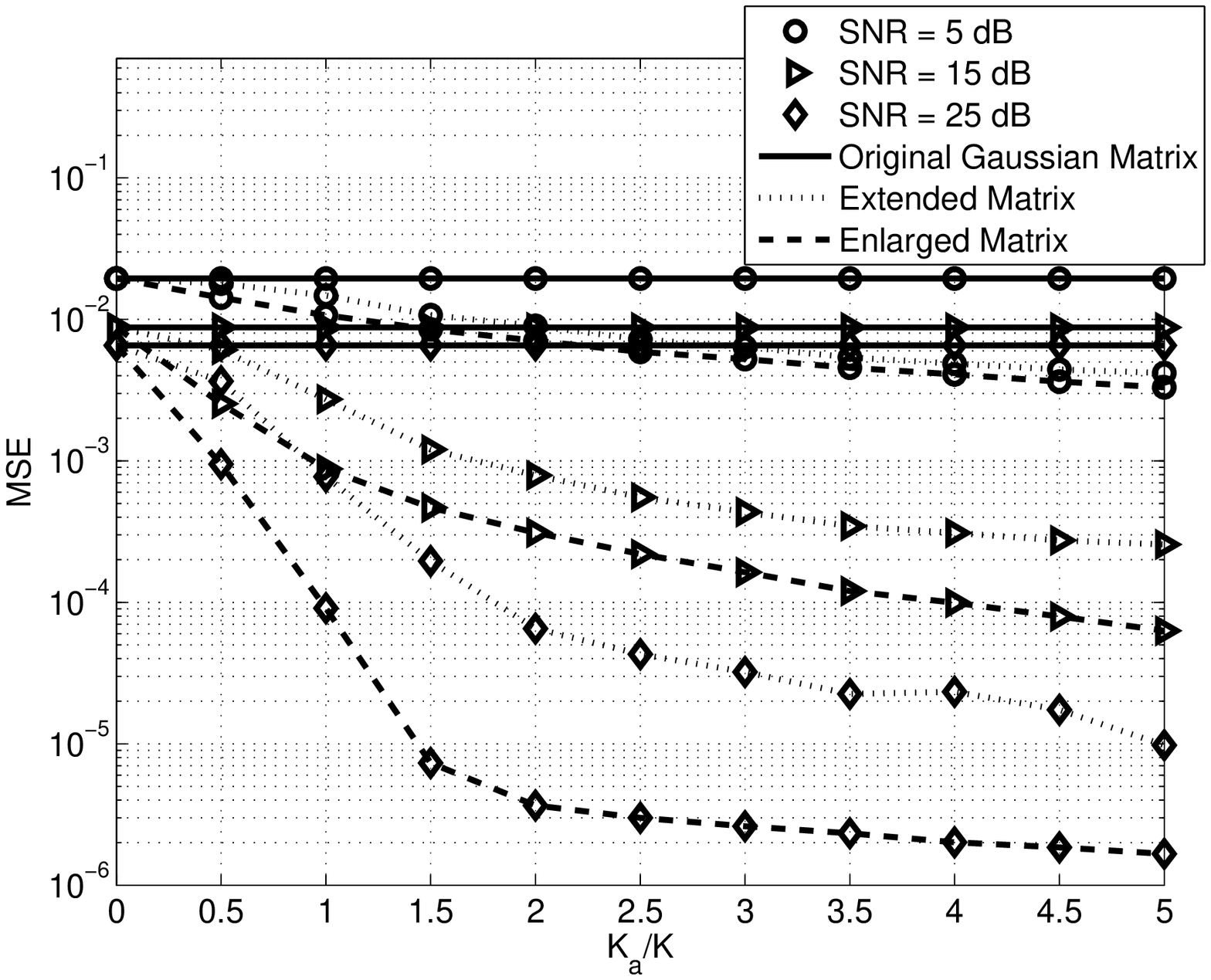}
\label{sfig:emprgaussianmsevska} } \subfigure[Measurement matrix with Bernoulli distributed entries] {
\includegraphics[angle=0,width=0.47\textwidth]{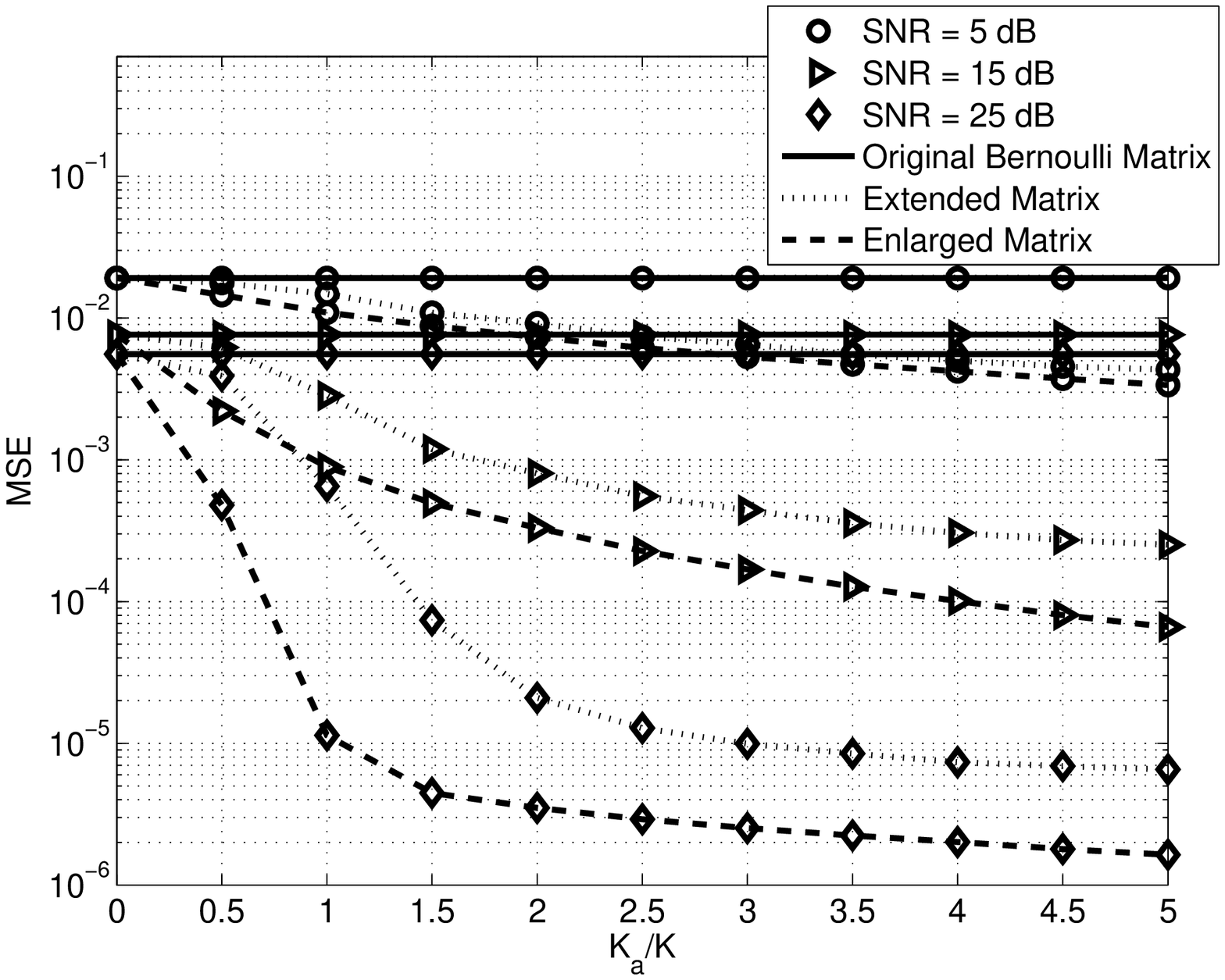}
\label{sfig:emprbernoullimsevska} }
\caption{Recovery based on the empirical risk minimization method: MSEs versus
$K_{a}/K$.} \label{fig:emprmsevska} \vspace{-0.3cm}
\end{figure}

Fig.~\ref{fig:emprmsevska} shows the MSEs obtained based on the
empirical risk minimization method for all three measurement
matrices versus the ratio $K_{a} / K$. The results are shown for
three different SNR values of 5, 15 and 25~dB. Two cases are
considered: (a)~the entries of the original and the enlarged
measurement matrices are generated as i.i.d. zero mean Gaussian
distributed random variables with variance $1/N$ and (b)~the
entries of the original and enlarged measurement matrices are
generated as i.i.d. zero mean Bernoulli distributed random
variables with variance as in case (a). The same conclusions as in
the first example can be drawn in this example. Moreover, the
results for cases (a) and (b) are also similar. Therefore, the
proposed segmented AIC indeed leads to significantly improved
signal recovery performance without increasing the number of BMIs.

\section{Conclusion} \label{sec:conclusion}
A new segmented compressed sampling method for AIC has been
proposed. According to this method, signal is segmented into $M$
segments and passed through $K$ BMIs of AIC to generate a $K
\times M$ matrix of sub-samples. Then, a number of correlated
samples larger than the number of BMIs is constructed by adding up
different subsets of sub-samples selected in a specific manner.
Due to the inherent structure of the method, the complexity of the
sampling device is almost unchanged, while the signal recovery
performance is shown to be significantly improved. The complexity
increase is only due to the $M$ times higher sampling rate and the
necessity to solve a larger size optimization problem at the
recovery stage, while the number of BMIs remains the same at the
sampling stage. The validity and superiority of the proposed
segmented AIC method over the conventional AIC is justified
through theoretical analysis of the RIP and the quality of signal
recovery. Simulation results also verify the effectiveness and
superiority of the proposed segmented AIC method and approve our
theoretical studies.

\section*{Appendix A: Proof of Theorem 1}
The total number of possible permutations of $\boldsymbol{z}$ is
$K!$. Let ${\cal{A}}$ be the set of permutations $\pi_{s},\; s =
1, \hdots, |{\cal A}|$ that satisfy the following condition
\begin{equation}
\pi_{s} (k) \neq \pi_{t} (k), \quad s \neq t, \; \forall s, t \in
\{1, \hdots, |{\cal A}| \}, \; \forall k \in \{1, \hdots, K\}.
\label{eq:allowedsetofperms}
\end{equation}
It is easy to see that the number of distinct permutations
satisfying the condition \eqref{eq:allowedsetofperms} is $K$, so
$|{\cal{A}}| = K$. It is also straightforward to see that the
choice of such $K$ distinct permutations is not unique. As a
specific choice, let the elements of ${\cal{A}}$, i.e., the
permutations $\pi_s, \; s=1, \hdots, K$, be
\begin{equation}
\pi_{s} (k) = \left((s+k-2) \; \text{mod} \; K \right) + 1, \quad
s, k = 1, \hdots, K \label{eq:oneofthechoicesforperms}
\end{equation}
with $\pi_{1}$ being the identity permutation, i.e., the
permutations that does not change $\boldsymbol{z}$.

Consider now the matrix $\boldsymbol{Z}$ which consists of $M$
columns $\boldsymbol{z}$. The $i$-th set of column permutations of
matrix $\boldsymbol{Z}$ is ${\cal P}^{(i)} = \{ \pi^{(i)}_{1},
\hdots, \pi^{(i)}_{M} \}$ and the corresponding permuted matrix is
$\boldsymbol{Z}^{{\cal P}^{(i)}}$. Let $\{ \pi^{(i)}_{1}, \hdots,
\pi^{(i)}_{M} \}$ be any combination of the $K$ permutations in
\eqref{eq:oneofthechoicesforperms}. Then there are $K^M$ possible
choices for ${\cal P}^{(i)}$. However, not all of these possible
choices are permissible by the conditions of the theorem.

Indeed, let the set ${\cal P}^{(1)}$ be a combination of
permutations from ${\cal A}$ that satisfies
\eqref{seq:requirementonpermi}. There are $I-1$ other sets ${\cal
P}^{(i)}, \; i=2, \hdots, I$ which satisfy both
\eqref{seq:requirementonpermi} and
\eqref{seq:requirementonpermiandj}. Gathering all such sets in one
set, we obtain the set ${\cal P} = \{ {\cal P}^{(1)}, \hdots,
{\cal P}^{(I)} \}$. Now let ${\cal P}^{(I+1)} = [\pi_{1}^{(I+1)},
\hdots, \pi_{M}^{(I+1)}]$ be one more set of permutations where
$\exists \pi_{m}^{(I+1)}, \; m = 1, \hdots, M$ such that
$\pi_{m}^{(I+1)} \notin {\cal A}$. An arbitrary $k$-th row of
$\boldsymbol{Z}^{{\cal P}^{(I+1)}}$ is $\left(
[\boldsymbol{Z}^{{\cal P}^{(I+1)}}]_{k,1}, \hdots,
[\boldsymbol{Z}^{{\cal P}^{(I+1)}}]_{k,M} \right)$ where
$[\boldsymbol{Z}^{{\cal P}^{(I+1)}}]_{k,1}, \hdots,
[\boldsymbol{Z}^{{\cal P}^{(I+1)}}]_{k,M} \in \{ 1, \hdots, K \}$.
This exact same row can be found as the first row of one of the
permuted matrices $\boldsymbol{Z}^{{\cal P}^{(i)}}$, ${\cal
P}^{(i)} \in {\cal P}$. Specifically, this is the permuted matrix
$\boldsymbol{Z}^{{\cal P}^{(i)}}$ that is obtained by applying the
permutations ${\cal P}^{(i)} = \left\{ \pi_{
[\boldsymbol{Z}^{{{\cal P}}^{(I+1)}}]_{k,1}}, \hdots, \pi_{
[\boldsymbol{Z}^{{{\cal P}}^{(I+1)}}]_{k,M}} \right\}$. The
permutations ${\cal P}^{(i)}$ either has to belong to ${\cal P}$
or being crossed out from ${\cal P}$ because of conflicting with
some other element ${\cal P}^{(l)} \in {\cal P}$, $l \neq i$. In
both cases, ${\cal P}^{(I+1)}$ can not be added to ${\cal P}$
because it will contradict the conditions
\eqref{seq:requirementonpermi} and
\eqref{seq:requirementonpermiandj}.

Therefore, the set ${\cal P}$ can be built using only the
permutations from the set ${\cal A}$, i.e., the $K$ permutations
in \eqref{eq:oneofthechoicesforperms}. Rearranging the rows of
$\boldsymbol{Z}^{{\cal P}^{(i)}}$ in a certain way, one can force
the elements in the first column of $\boldsymbol{Z}^{{\cal
P}^{(i)}}$ to appear in the original increasing order, i.e.,
enforce the first column be equivalent to the vector of indexes
$\boldsymbol{z}$. It can be done by applying to each permutation
in the set ${\cal P}^{(i)}$ the inverse permutation $\left(
{\pi^{(i)}_{1}} \right)^{-1}$, which itself is one of the
permutations in \eqref{eq:oneofthechoicesforperms}. Therefore, the
set ${\cal P}^{(i)} = \{ \pi^{(i)}_{1}, \hdots, \pi^{(i)}_{M} \}$
can be replaced by the equivalent set $\left\{ \left(
{\pi^{(i)}_{1}} \right)^{-1} \pi^{(i)}_{1}, \hdots, \left(
{\pi^{(i)}_{1}} \right)^{-1} \pi^{(i)}_{M} \right\} = \left\{
\pi_{1}, \hdots, \left( {\pi^{(i)}_{1}} \right)^{-1} \pi^{(i)}_{M}
\right\}$, where $\pi_{1}$ is the identity permutation and $\left(
{\pi^{(i)}_{1}} \right)^{-1} \pi^{(i)}_{j} \in {\cal A}$. Hence,
we can consider only the permutations of the form ${\cal P}^{(i)}
= \{ \pi_{1}, \hdots, \pi_j^{(i)}, \hdots, \pi^{(i)}_{M} \}$.
Since the condition \eqref{seq:requirementonpermi} requires that
$\pi^{(i)}_{2}$ should be different from $\pi_{1}$, the only
available options for the permutations on the second column of
$\boldsymbol{Z}$ are the $K-1$ permutations $\pi_{2}, \hdots,
\pi_{K}$ in \eqref{eq:oneofthechoicesforperms}. Therefore, $I$ at
most equals $K-1$. Note that $I$ can be smaller than  $K-1$ if for
some $i \in \{ 1, \hdots, K-1 \}$, \; $K/gcd(i,K) < M$ (also see
Example~1 after Theorem~1). Thus, in general $I \leq K-1$.

\section*{Appendix B: Proof of Lemma 1}
Let all the rows of $\left(\boldsymbol{\Phi}_{e}\right)_{\cal T}$
be partitioned into two sets of sizes (cardinality) as close as
possible to each other, where all elements in each set are
guaranteed to be statistically independent. In particular, note
that the elements of the new $K_a$ rows of $\boldsymbol{\Phi}_e$
are chosen either from the first $K_{a} + M - 1$ rows of
$\boldsymbol{\Phi}$ if $K_{a} + M - 1 < K$ or from the whole
matrix $\boldsymbol{\Phi}$. Therefore, if $K_{a} + M - 1 < K$, the
last $K - K_{a} - M + 1$ rows of $\boldsymbol{\Phi}$ play no role
whatsoever in the process of extending the measurement matrix and
they are independent on the rows of $\boldsymbol{\Phi}_1$ in
\eqref{eq:newmeasurementmatrix}. These rows are called unused
rows. Thus, one can freely add any number of such unused rows to
the set of rows in $\boldsymbol{\Phi}_1$ without disrupting its
status of being formed by independent Gaussian variables. Since
$\text{min} \{ K, K_{a} + M - 1 \} \leq \lceil \left( K + K_{a}
\right) / 2 \rceil$, there exist at least $\lfloor \left( K +
K_{a} \right) / 2 \rfloor - K_a$ unused rows which can be added to
the set of rows in $\boldsymbol{\Phi}_1$. Such process describes
how the rows of $\left( \boldsymbol{\Phi}_{e} \right)_{\cal T}$
are split into the desired sets $\left( \boldsymbol{\Phi}_{e}
\right)_{\cal T}^1$ and $\left( \boldsymbol{\Phi}_{e}
\right)_{\cal T}^2$ of statistically independent elements. As a
result, the first matrix $\left( \boldsymbol{\Phi}_{e}
\right)_{\cal T}^1$ includes the first $\lceil \left( K + K_{a}
\right) / 2 \rceil$ rows of $\left(
\boldsymbol{\Phi}_{e}\right)_{\cal T}$, while the rest of the rows
are included in $\left( \boldsymbol{\Phi}_{e} \right)_{\cal T}^2$.

Since the elements of the matrices $( \boldsymbol{\Phi}_{e}
)_{\cal T}^1$ and $(\boldsymbol{\Phi}_{e} )_{\cal T}^2$ are i.i.d.
Gaussian, they will satisfy (\ref{eq:ripconstant}) with
probabilities equal or larger than $1 - 2 \left( 12 /
\delta_{S} \right)^S e^{-C_{0} \lceil K_{e} / 2 \rceil}$ and $1 -
2 \left( 12 / \delta_{S} \right)^S e^{-C_{0} \lfloor K_{e} / 2
\rfloor}$, respectively. Therefore, both matrices $(
\boldsymbol{\Phi}_{e} )_{\cal T}^1$ and $ ( \boldsymbol{\Phi}_{e}
)_{\cal T}^2$ satisfy (\ref{eq:ripconstant}) simultaneously with
the common probability
\begin{equation}
\text{Pr} \{ (\boldsymbol{\Phi}_{e} )_{\cal T}^i \;
\text{satisfies (\ref{eq:ripconstant})}\} \geq 1 - 2 (12 /
\delta_{S} )^S e^{-C_{0} \lfloor K_{e} / 2 \rfloor}, \quad i = 1,
2. \label{eq:ripsatisfactionprob}
\end{equation}

Let $K_{1}^{\prime} \triangleq \lceil K_{e} / 2 \rceil$ and
$K_{2}^{\prime} \triangleq \lfloor K_{e} / 2 \rfloor$. Consider
the event when both $(\boldsymbol{\Phi}_{e} )_{\cal T}^1$ and
$(\boldsymbol{\Phi}_{e} )_{\cal T}^2$ satisfy
(\ref{eq:ripconstant}). Then the following inequality hold for any
vector $\boldsymbol{c} \in \mathbb{R}^S$:
\begin{equation}
\sum_{i=1}^2 { \frac{K_{i}^{\prime}}{N} ( 1 - \delta_S ) \|
\boldsymbol{c} \|_{l_{2}}^2} \leq \sum_{i=1}^2 { \| (
\boldsymbol{\Phi}_{e} )_{\cal T}^{i} \boldsymbol{c} \|_{l_{2}}^2}
\leq \sum_{i=1}^2 { \frac{K_{i}^{\prime}}{N} ( 1 + \delta_S ) \|
\boldsymbol{c} \|_{l_{2}}^2} \label{eq:jointeventleadstorip1}
\end{equation}
or, equivalently,
\begin{align}
\frac{K_e}{N} ( 1 - \delta_S ) \| \boldsymbol{c} \|_{l_{2}}^2 \leq
\| ( \boldsymbol{\Phi}_{e} )_{\cal T} \boldsymbol{c}\|_{l_{2}}^2
\leq \frac{K_e}{N} ( 1 + \delta_S ) \| \boldsymbol{c}
\|_{l_{2}}^2. \label{eq:jointeventleadstorip2}
\end{align}
Therefore, if both matrices $( \boldsymbol{\Phi}_{e} )_{\cal T}^1$
and $(\boldsymbol{\Phi}_{e} )_{\cal T}^2$ satisfy
(\ref{eq:ripconstant}), then the matrix $( \boldsymbol{\Phi}_{e}
)_{\cal T}$ also satisfies (\ref{eq:ripconstant}). Moreover, the
probability that $( \boldsymbol{\Phi}_{e} )_{\cal T}$ does not
satisfy (\ref{eq:ripconstant}) can be found as
\begin{eqnarray}
\text{Pr} \{ ( \boldsymbol{\Phi}_{e} )_{\cal T} \;\text{does not
satisfy (\ref{eq:ripconstant})}\} \!\!&\leq&\!\! \text{Pr} \{ (
\boldsymbol{\Phi}_{e} )_{\cal T}^1 \;\text{or} \; (
\boldsymbol{\Phi}_{e} )_{\cal T}^2 \; \text{does
not satisfy (\ref{eq:ripconstant})}\} \nonumber \\
\!\!&\stackrel{(a)}{\leq}&\!\! \sum_{i=1}^2 {\text{Pr} \{ (
\boldsymbol{\Phi}_{e} )_{\cal T}^i \; \text{does
not satisfy (\ref{eq:ripconstant})}\}} \nonumber \\
\!\!&\stackrel{(b)}{\leq}&\!\! 4 \left( 12 / \delta_{S} \right)^S
e^{-C_{0} \lfloor K_{e} / 2 \rfloor} \label{eq:totripprob}
\end{eqnarray}
where the inequality (a) follows from the union bounding and the
inequality (b) follows from (\ref{eq:ripsatisfactionprob}). Thus,
the inequality (\ref{eq:phiripsatisfactionprob}) holds.

\section*{Appendix C: Proof of Theorem~2}
According to (\ref{eq:phiripsatisfactionprob}), the matrix $\left(
\boldsymbol{\Phi}_{e} \right)_{\cal T}$ does not satisfy
(\ref{eq:ripconstant}) with probability less than or equal to $4
\left( 12 / \delta_{S} \right)^S e^{-C_{0} \lfloor K_{e} / 2
\rfloor}$ for any subset ${\cal T} \subset \{1, \hdots, N\}$ of
cardinality $S$. Since there are ${N \choose S} \leq (Ne/S)^S$
different subsets ${\cal T}$ of cardinality $S$,
$\boldsymbol{\Phi}_e$ does not satisfy the RIP with probability
\begin{eqnarray}
&&\!\!\!\!\!\!\!\!\!\!\!\!\!\!\!\!\! \text{Pr}
\{\boldsymbol{\Phi}_{e} \; \text{does not satisfy RIP} \} \leq 4
{N \choose S} \left( 12 / \delta_{S} \right)^S e^{-C_{0}
\lfloor K_{e} / 2 \rfloor} \nonumber \\
&\leq&\!\!\! 4 \left( Ne/S \right)^S \left( 12 / \delta_{S}
\right)^S e^{-C_{0} \lfloor K_{e} / 2 \rfloor} = 4 e^{-\left(C_{0}
\lfloor K_{e} / 2 \rfloor - S \left[\log \left( Ne/S \right) + \log
\left(12/\delta_S\right) \right] \right)} \nonumber \\
&\leq&\!\!\! 4 e^{-\left\{ C_{0} \lfloor K_{e}/2 \rfloor - C_{3}
\left[ \log \left(Ne/S\right) + \log \left( 12 / \delta_S \right)
\right] \lfloor K_{e}/2 \rfloor / \log \left( N/S \right) \right
\}} \nonumber \\
&=&\!\!\! 4 e^{-\left\{C_{0} - C_{3} \left[ 1 + \left( 1 + \log
\left( 12 / \delta_S \right) \right) / \log \left( N/S \right)
\right] \right\} \lfloor K_{e} / 2 \rfloor}.
\label{eq:ripprobequation1}
\end{eqnarray}
Setting $C_{4}=C_{0} - C_{3} \left[ 1 + \left( 1 + \log \left( 12 /
\delta_S \right) \right) / \log \left( N/S \right) \right]$ and
choosing $C_3$ small enough that guarantees that $C_4$ is
positive, we obtain (\ref{finprob}).

\section*{Appendix D: Proof of Lemma 2}
The method of the proof is the same as the one used to prove
Lemma~\ref{lem:lemripprob} and is based on splitting the rows of
$\boldsymbol{\Phi}_{e}$ into a number of sets with independent
entries. Here, the splitting is carried out as shown in
(\ref{eq:extendedmeasurementmatrixgeneral}).

Let $( \boldsymbol{\Phi}_{e} )_{\cal T}^i, \; i=1, \hdots, n_{p} -
1$ be the matrix containing the $(i-1) K + 1$-th to the $iK$-th
rows of $( \boldsymbol{\Phi}_{e} )_{\cal T}$. The last $K_{e} -
(n_{p}-1) K$ rows of $( \boldsymbol{\Phi}_{e} )_{\cal T}$ form the
matrix $( \boldsymbol{\Phi}_{e} )_{\cal T}^{n_{p}}$. Since the
matrices $( \boldsymbol{\Phi}_{e} )_{\cal T}^{i}, \; i=1, \hdots,
n_{p} -  1$ consist of independent entries, they satisfy
(\ref{eq:ripconstant}) each with probability of at least $1 - 2
\left( 12 / \delta_{S} \right)^S e^{-C_{0} K}$. For the same
reason, the matrix $( \boldsymbol{\Phi}_{e} )_{\cal T}^{n_p}$
satisfies (\ref{eq:ripconstant}) with probability greater than or
equal to $1 - 2 \left( 12 / \delta_{S} \right)^S e^{-C_{0}
K_{n_p}}$. In the event that all the matrices $(
\boldsymbol{\Phi}_{e} )_{\cal T}^{i}, \; i=1,..,n_{p}$ satisfy
(\ref{eq:ripconstant}) simultaneously, for $\boldsymbol{c} \in
\mathbb{R}^S$ we have
\begin{eqnarray}
\sum_{i=1}^{n_p} {\frac{K_{i}}{N} ( 1 - \delta_S ) \|
\boldsymbol{c} \|_{l_{2}}^2} \!\!\!\!&\leq&\!\!\!\!
\sum_{i=1}^{n_p} { \| ( \boldsymbol{\Phi}_{e} )_{\cal T}^{i}
\boldsymbol{c} \|_{l_{2}}^2} \leq \sum_{i=1}^{n_p} {\frac{K_{i}}
{N}
( 1 + \delta_S) \| \boldsymbol{c} \|_{l_{2}}^2} \nonumber \\
\Rightarrow && \frac{K_e}{N}( 1 -\delta_S ) \|\boldsymbol{c}
\|_{l_{2}}^2 \leq \| ( \boldsymbol{\Phi}_{e} )_{\cal T}
\boldsymbol{c} \|_{l_{2}}^2 \leq \frac{K_e}{N}( 1 + \delta_S ) \|
\boldsymbol{c} \|_{l_{2}}^2.
\label{eq:jointeventleadstoripgeneral1}
\end{eqnarray}

Therefore, using the union bound and
\eqref{eq:jointeventleadstoripgeneral1}, we can conclude that
\begin{eqnarray}
& &\!\!\!\!\!\!\!\!\!\!\!\!\!\!\!\!\!\!\! \text{Pr} \{ (
\boldsymbol{\Phi}_{e} )_{\cal T} \; \text{does not satisfy
(\ref{eq:ripconstant})} \} \leq \sum_{i=1}^{n_p} {\text{Pr} \{ (
\boldsymbol{\Phi}_{e} )_{\cal T}^i \; \text{does
not satisfy (\ref{eq:ripconstant})} \}} \nonumber \\
&\leq&\!\! 2 (n_{p} - 1) \left( 12 / \delta_{S} \right)^S \left(
e^{-C_{0} K} \right) + 2\left( 12 / \delta_{S} \right)^S \left(
e^{-C_{0} K_{n_{p}}} \right) \label{eq:totripprobgeneral}
\end{eqnarray}
which proves the lemma.

\section*{Appendix E: Proof of Theorem 3}
According to Lemma~2, for any subset ${\cal T} \subset
\{1,\hdots,N\}$ of cardinality $S$, the probability that $(
\boldsymbol{\Phi}_{e} )_{\cal T}$ does not satisfy
(\ref{eq:ripconstant}) is less than or equal to $2 (n_{p} - 1)
\left( 12 / \delta_{S} \right)^S \left( e^{-C_{0} K} \right) +
2\left( 12 / \delta_{S} \right)^S \left( e^{-C_{0} K_{n_{p}}}
\right)$. Using the fact that there are ${N \choose S} \leq
(Ne/S)^S$ different subsets $\cal T$, the probability that the
extended measurement matrix $\boldsymbol{\Phi}_e$ does not satisfy
the RIP can be computed as
\begin{eqnarray}
&&\!\!\!\!\!\!\!\!\!\!\!\!\!\!\!\!\! \text{Pr} \{
\boldsymbol{\Phi}_{e} \; \text{does not satisfy the RIP}\}
\nonumber \leq 2 ( n_{p} - 1 ) {N \choose S} \left( 12 /
\delta_{S} \right)^S e^{-C_{0} K} + 2 {N \choose S} \left(
12 / \delta_{S} \right)^S e^{-C_{0} K_{n_p}} \nonumber \\
&\leq&\!\!\! 2 (n_{p} - 1 ) \left( Ne/S \right)^S
\left(12/\delta_{S} \right)^S e^{-C_{0}K} + 2 \left( Ne/S
\right)^S \left( 12 / \delta_{S} \right)^S e^{-C_{0}
K_{n_p}} \nonumber \\
&=&\!\!\! 2 ( n_{p} - 1 ) e^{-\left( C_{0} K -S \left[ \log \left(
Ne/S \right) + \log \left( 12 / \delta_S \right) \right] \right)} +
2 e^{-\left( C_{0} K_{n_p} -S \left[ \log \left( Ne/S \right) + \log
\left( 12 / \delta_S \right)
\right] \right) }\nonumber \\
&\leq&\!\!\! 2 (n_{p} - 1) e^{-\left\{ C_{0} K - \frac{C_{3}
K_{n_p}} {K} \left[ \log \left( Ne/S \right) + \log \left( 12 /
\delta_S \right) \right] K / \log \left ( N / S \right) \right\}}
\nonumber \\
&+&\!\!\! 2 e^{-\left\{ C_{0} K_{n_p} - C_{3} K_{n_p} \left[ \log
\left( Ne/S \right) + \log \left( 12 / \delta_S \right) \right]
K_{n_p} / \log
\left( N/S \right) \right\} } \nonumber \\
&=&\!\!\! 2(n_{p} \!-\! 1) e^{-\left\{C_{0} -\frac{C_{3}
K_{n_p}}{K} \left[1 + \left( 1 + \log \left( 12 / \delta_S \right)
\right) / \log \left( N/S \right) \right] \right\} K} +
2e^{-\left\{C_{0} -C_{3} \left[ 1 + \left( 1 + \log \left( 12 /
\delta_S \right) \right) / \log \left( N/S \right) \right]
\right\} K_{n_p}}. \nonumber \\ \label{eq:generalripprobequation1}
\end{eqnarray}
Denoting the constant terms as $C_{4} = C_{0} - C_{3} \left[ 1 +
\left( 1 + \log \left( 12 / \delta_S \right) \right) / \log \left(
N/S \right) \right]$ and $C_{4}^{\prime} = C_{0} - (C_{3} K_{n_p}
/ K)$ $\times \left[ 1 + \left( 1 + \log \left( 12 / \delta_S
\right) \right) / \log \left( N/S \right) \right]$, and choosing
$C_3$ small enough in order to guarantee that $C_4$ and
$C_{4}^{\prime}$ are positive, we obtain
(\ref{eq:finprobgeneral}).


\begin{thebibliography}{11}
\bibitem{CandesWakin08}
E.~J. Candes, and M.~B. Wakin,
\newblock ``An introduction to compressive sampling,''
\newblock {\em IEEE Signal Processing Magazine}, vol. 25, pp.~21--30, March
  2008.

\bibitem{CandesTao05}
E.~Candes and T.~Tao,
\newblock ``Decoding by linear programming,''
\newblock {\em IEEE Trans. Inf. Theory}, vol. 51, pp. 4203--4215, Dec. 2005.

\bibitem{Donoho06b}
D.~Donoho,
\newblock ``Compressed sensing,''
\newblock {\em IEEE Trans. Inf. Theory}, vol. 52, pp. 1289--1306, Apr. 2006.

\bibitem{HauptNowak06}
J.~Haupt and R.~Nowak,
\newblock ``Signal reconstruction from noisy random projections,''
\newblock {\em IEEE Trans. Inf. Theory}, vol. 52, pp.~4036--4048, Sept. 2006.

\bibitem{Wakinetal06}
M.~Wakin, J.~N. Laska, M.F. Duarte, D.~Baron, S.~Sarvotham,
D.~Takhar, K.F.  Kelly, and R.G. Baraniuk,
\newblock ``An architecture for compressive imaging,''
\newblock in {\em Proc. IEEE ICIP}, Atlanta, USA, Oct. 2006, pp.~1273--1276.

\bibitem{Bajwaetal07a}
W.~Bajwa, J.~Haupt, A.~Sayeed, and R.~Nowak,
\newblock ``Joint source–channel communication for distributed estimation in sensor networks,''
\newblock {\em IEEE Trans. Inf. Theory}, vol. 53, pp. 3629--3653, Oct. 2007.

\bibitem{YuHoyosSadler08}
Z.~Yu, S.~Hoyos, and B.~M. Sadler,
\newblock ``Mixed-signal parallel compressed sensing and reception for cognitive radio,''
\newblock in {\em Proc. IEEE ICASSP}, Las Vegas, USA, Apr. 2008, pp.
  3861--3864.


\bibitem{TaubockHlawatsch08}
G.~Taubock, and F.~Hlawatsch,
\newblock ``A compressed sensing technique for OFDM channel
estimation in mobile environments: Exploiting channel sparsity for reducing pilots,''
\newblock in {\em Proc. IEEE ICASSP}, Las Vegas, USA, Apr. 2008, pp. 2885--2888.

\bibitem{Bajwaetal08}
W.~U.~Bajwa, J.~Haupt, G.~Raz, and R.~Nowak,
\newblock ``Compressed channel sensing,''
\newblock in {\em Proc. IEEE CISS}, Princeton, USA, Mar. 2008, pp. 5--10.



\bibitem{Eldar09} Y.~C.~Eldar,
\newblock ``Compressed sensing of analog signals in shift-invariant spaces,''
\newblock {\em IEEE Trans. Sig. Processing}, vol. 57, No. 8, pp.~2986--2997, Aug. 2009.

\bibitem{MishaliEldar09}
M.~Mishali and Y.~C.~Eldar,
\newblock ``Blind multiband signal reconstruction: Compressed sensing for analog signals,''
\newblock {\em IEEE Trans. Sig. Processing}, vol. 57, No. 3, pp.~993--1009, Mar. 2009.

\bibitem{deBoor} C.~de~Boor, R.~De~Vore, and A.~Ron,
\newblock ``The structure of finite genetated shift-invariant spaces in $L_2 (\mathbb{R}^d)$,''
\newblock {\em J. Funct. Anal.}, vol. 119, No. 1, pp.~37--78, 1994.

\bibitem{LuDo}
Y.~M.~Lu and M.~N.~Do,
\newblock ``A theory for sampling signals from a union of subspaces,''
\newblock {\em IEEE Trans. Sig. Processing}, vol. 56, No. 6, pp.~2334--2345, Jun. 2008.

\bibitem{EldarMishali09} Y.~C.~Eldar and M.~Mishali,
\newblock ``Robust recovery of signals from a structured union of subspaces,''
\newblock {\em IEEE Trans. Inf. Theory}, vol. 55, No. 11, pp.~5302--5316, Nov. 2009.

\bibitem{Laskaetal08}
J.~N. Laska, S.~Kirolos, M.F. Duarte, T.S. Ragheb, R.G. Baraniuk,
and Y.~Massoud,
\newblock ``Theory and implementation of an analog-to-information converter
  using random demodulation,''
\newblock in {\em Proc. IEEE ISCAS}, New Orleans, USA, May 2007, pp.~1959--1962.

\bibitem{OmidSergiy}
O.~Taheri and S.~A.~Vorobyov,
\newblock ``Segmented compressed sampling for analog-to-information
conversion,''
\newblock in {\em Proc. IEEE CAMSAP}, Aruba, Dutch Antilles, Dec. 2010, pp.~113--116.

\bibitem{Bajwaetal07b}
W.~Badjwa, J.~D. Haupt, G.~M. Raz, S.~J. Wright, and R.~D. Nowak,
\newblock ``Toeplitz-structured compressed sensing matrices,''
\newblock in {\em Proc. IEEE SSP}, Madison, USA, Aug. 2007, pp. 294--298.

\bibitem{BarStee07}
R.~Baraniuk and P.~Steeghs,
\newblock ``Compressive radar imaging,''
\newblock in {\em Proc. IEEE Radar Conf.}, Waltham, MA, USA, Apr.~2007.

\bibitem{DonohoTanner09}
D.~L.~Donoho and J.~Tanner,
\newblock ``Counting faces of randomly projected polytopes when the projection radically lowers dimension,''
\newblock {\em Journal of the American Math. Society}, vol. 22, no. 1, pp. 1--53, Jan. 2009.

\bibitem{CandesTao06}
E.~Candes and T.~Tao,
\newblock ``Near optimal signal recovery from random projections: universal encoding stategies?,''
\newblock {\em IEEE Trans. Inf. Theory}, vol.~52, No.~12 , pp.~5406--5425, Dec. 2006.

\bibitem{Baraniuketal08}
R.~Baraniuk, M.~Davenport, R.~De Vore, and M.~Wakin,
\newblock ``A simple proof of the restricted isometry property for random
  matrices,''
\newblock {\em Constructive Approximation}, Jan. 2008.

\bibitem{Donoho06a}
D.~Donoho,
\newblock ``For most large underdetermined systems of linear equations the
  minimal $l_1$-norm solution is also the sparsest solution,''
\newblock {\em Communi. Pure and Applied Math.}, vol. 59, pp.~797--829, Jun.
  2006.

\bibitem{CandesRombergTao05}
E.~Candes, J.~Romberg, and T.~Tao,
\newblock ``Stable signal recovery from incomplete and inaccurate
  measurements,''
\newblock {\em Communi. Pure and Applied Math.}, vol. 59, pp. 1207--1223, Aug.
  2006.

\bibitem{Vapnik98}
V. ~N.~Vapnik,
\newblock {\em Statistical Learning Theory},
\newblock Wiley, New York, 1998.

\bibitem{AngelosanteGiannakis09}
D.~Angelosante, G.~B.~Giannakis,
\newblock ``RLS-weighted LASSO for adaptive estimation of sparse signals,''
\newblock {\em Proc. IEEE ICASSP}, Taipei, Taiwan, Apr. 2009, pp. 3245--3248.

\bibitem{Craig33}
C.~Craig,
\newblock ``On the Tchebycheff inequality of Bernstein,''
\newblock {\em Ann. Math. Stat.}, vol. 4, no. 2, pp. 94--102, May 1933.

\bibitem{YinZhang}
Y.~Zhang,
\newblock ``Theory of compressive sensing via $l_1$-minimization: A
non-RIP analysis and extensions,''
\newblock {\em }, 2010.

\end{thebibliography}
\end{document}